\documentclass[journal,10pt]{IEEEtran}
\usepackage[utf8]{inputenc}
\usepackage[T1]{fontenc}
\usepackage{multirow}
\usepackage[table]{xcolor}
\usepackage{graphicx}
\usepackage{wrapfig}
\usepackage{pgf}
\usepackage{soul}

\usepackage{algpseudocode}
\usepackage{amsmath}
\usepackage{amsfonts}
\usepackage{mathtools}
\usepackage{amssymb}
\usepackage{mathrsfs}
\usepackage{cite}
\usepackage{subfigure}
\usepackage[ruled,vlined,linesnumbered]{algorithm2e}
\usepackage{tikz}
\usetikzlibrary{shapes.geometric, arrows}
\usepackage{tablefootnote}
\everymath{\displaystyle}

\newcommand\drg{\textcolor{red}}

\tikzstyle{startstop} = [rectangle, rounded corners, minimum width=2.5cm, minimum height=1cm,text centered, text width=3cm, draw=black, font=\fontsize{9pt}{9pt}\selectfont]
\tikzstyle{io} = [trapezium, trapezium left angle=70, trapezium right angle=110, minimum width=1cm, minimum height=1cm, text centered, draw=black, font=\fontsize{9pt}{9pt}\selectfont]
\tikzstyle{process} = [rectangle, minimum width=2.5cm, minimum height=1cm, text width=3cm, draw=black, font=\fontsize{9pt}{9pt}\selectfont]
\tikzstyle{decision} = [diamond, aspect = 2, minimum width=2cm, minimum height=1cm, text centered, text width=2cm, draw=black, font=\fontsize{9pt}{9pt}\selectfont]
\tikzstyle{arrow} = [thick,->,>=stealth]

\begin{document}

\title{{Energy-Efficient Wake-Up Signalling for Machine-Type %communications 
Devices % 
Based on Traffic-Aware Long-Short Term Memory Prediction}} %in Beyond 5G Systems}

\author{{David E. Ruíz-Guirola, Carlos A. Rodríguez-López, Samuel Montejo-Sánchez, \IEEEmembership{Senior Member, IEEE}, \\Richard Demo Souza, \IEEEmembership{Senior Member, IEEE}, Onel L. A. López, \IEEEmembership{Member, IEEE}, and Hirley Alves, \IEEEmembership{Member, IEEE}}

\thanks{
David E. Ruíz-Guirola, Onel L. A. López and Hirley Alves are with the Centre for Wireless Communications University of Oulu, Finland. \{{David.RuizGuirola, Onel.AlcarazLopez, Hirley.Alves}@oulu.fi\} 
Carlos A. Rodríguez-López is with the {Department of Electronics and Telecommunications, Universidad Central ``Marta Abreu'' de Las Villas}, Santa Clara, Cuba. \{crodrigz@uclv.edu.cu\}
Samuel Montejo-Sánchez is with the {Programa Institucional de Fomento a la Investigación, Desarrollo e Innovación, Universidad Tecnológica Metropolitana}, Santiago, Chile. \{smontejo@utem.cl\}
Richard Demo Souzais with the Federal University of Santa Catarina, Florianópolis, SC, Brazil. \{richard.demo@ufsc.br\}

This work has been partially supported in Chile by ANID FONDECYT Iniciaci\'on No. 11200659, %SCC-PIDi-UTEM 
FONDEQUIP-EQM180180, and Collaborative Research Activities between PIDi/UTEM and FIE/UCLV, in Brazil by CNPq (402378/2021-0, 305021/2021-4), Print CAPES-UFSC ``Automation 4.0'', and RNP/MCTIC (Grant 01245.010604/2020-14) 6G Mobile Communications Systems, and in Finland by 6Genesis Flagship (Grant no. 318927) and Tekniikan Edistämissäätiön.
}
}

\maketitle

\begin{abstract}

Reducing energy consumption is a pressing issue in low-power machine-type communication (MTC) networks. {In this regard, the Wake-up Signal (WuS) technology, which aims to minimize the energy consumed by the radio interface of the machine-type devices (MTDs), stands as a promising solution.
However,} state-of-the-art WuS mechanisms use static operational parameters{, so they cannot efficiently adapt to the system dynamics.} To overcome this, we design a simple but efficient neural network to predict MTC traffic patterns and configure WuS accordingly{. Our proposed} forecasting WuS {(FWuS) leverages} an accurate long-short term memory (LSTM)-based traffic prediction that allows extending the sleep time {of MTDs} by avoiding frequent page monitoring occasions in idle state.
Simulation results show the effectiveness of %both the predictor and the proposed WuS configuration. 
our {approach.}  
The traffic prediction errors are shown to be below 4\%, being false alarm and miss-detection probabilities respectively below 8.8\% and 1.3\%. 
{In terms of energy consumption reduction, FWuS can outperform the best benchmark mechanism in up to 32\%.} 
Finally, we certify the ability {of
FWuS
to} dynamically adapt to traffic density changes{, promoting low-power MTC scalability.}

\end{abstract} 


\section{Introduction}\label{sec1}
%
%The fifth generation (5G) and beyond 
Future cellular networks need to support and drive a large variety of existing, emerging, and %/or 
even unforeseen, Internet of Things (IoT) use cases%, including machine type communications (MTC)
{~\cite{mahmood2021machine,9057670,xia2019emerging}.} 
Indeed, machine-type communication {(MTC)
%, also regarded as machine-to-machine (M2M) communication, 
is} an essential component of {the} fifth generation (5G) {of} wireless networks { as it enables machine-type devices (MTDs) to communicate and interact with each other without human interventions}
%In fact, massive MTC (mMTC), which focuses on providing connectivity to a huge number of MTC devices (MTDs), is a key service mode in the fifth generation (5G) of cellular systems~
\cite{akpakwu2017survey}. 
%Unlike traditional user \textcolor{blue}{equipments}, machine-type devices (MTDs) can communicate with each other autonomously without human interventions. 
%In our daily lives, more and more machines equipped with communication modules act as  MTDs, thus, enabling a broad range of applications
{Applications include}
mission critical services, intelligent transportation {systems, fleet} management, smart grid, industrial automation, real-time monitoring/control, and remote medical systems~\cite{manogaran2020response, YAN2022102601, delay, zanella2014internet, bayindir2016path, ratasuk2015recent}. 
%Fueled by the booming IoT market, 
{Consequently, the} number of MTDs is explosively growing, and billions of MTDs, such as sensors, actuators, and meters, are predicted to be operational in the coming years. 

%\textcolor{blue}{Most MTC deployments require} ubiquitous wireless connectivity services and energy efficient techniques to prolong MTDs' lifetime, which poses unprecedented challenges on the radio access network~\cite{9375479}. \textcolor{blue}{Indeed,

%Although there has been important advances, 
Energy efficiency is a key design requirement for IoT networks composed of {MTDs
that must operate for several years without battery recharging or  replacement~\cite{Adrx,zhou2019online}. Most of these MTDs have} 
limited energy resources due to their small size, low cost and/or hard-to-reach locations~\cite{li2020power,benbuk2021tunable,shehab2020traffic}{, which poses unprecedented challenges on the radio access network~\cite{9375479}.} 
%{Indeed, MTDs are required to operate for several years without recharging or battery replacement~\cite{Adrx}.} 
%In these applications,
%\textchange{In this context, accurate traffic prediction would be critical for efficient green networks~\cite{graph, canweachieve, hybrid_deep}.} 
%Therefore, energy efficiency improvement, especially for MTDs, is of great concern and interest for both %the academic and industrial communities
%academy and industry~\cite{zhou2019online}. 
A fundamental approach lies in bringing the complexity of IoT devices down by decreasing the computation capability, using low-order modulations, and introducing deep sleep modes~\cite{dian2020lte}. 
{Indeed, idle channel monitoring  is one of the major energy consumption sources at the {MTDs}, particularly in networks with low traffic~\cite{mazloum2020interference}. In this regard,} the 3rd generation partnership project (3GPP)~introduced wake up radios (WuRs) in its Release 15~\cite{3gpp2019ts38} to prolong the lifetime of battery-powered devices. For both, long term evolution for MTC (LTE-M) and narrow-band IoT (NB-IoT), device power consumption in idle mode is reduced by exploiting a wake-up signal (WuS). {Similarly, IEEE 802.11 introduced WuS in the IEEE 802.11ba amendment~\cite{WuS,stepanova2020joint}.
%,  aimed at significantly reducing the energy consumption of client devices~
}

{For many NB-IoT use cases envisioned by 3GPP, communication is infrequent over long periods of time~\cite{odelberg20212}, and therefore the standard includes Discontinuous or Extended Discontinuous Receive (DRX or eDRX) mode in idle operation~\cite{odelberg20212}. {During DRX/eDRX, the device waits for the reception of regular paging events (occasions) to decide whether to switch to active mode. Such paging events are preceded by the transmission of a narrowband WuS} %, In narrowband WuS prepends these paging events with a unique correlation-based orthogonal frequency-division multiplexing (OFDM) Zadoff-Chu sequence 
to allow the device to remain in a low power state by default, and only wake-up to decode a paging event if the WuS is identified~\cite{3gpp}. %For example, an NB-IoT application might expect availability of NB-IoT WuS.
}

WuS is transmitted within a configurable time before the device is paged, {page occasion (PO), maximizing the} sleep time~\cite{9626155, heins2022nb, mazloum2020interference}. % As MTDs avoid frequent wake-ups for listening paging information that is not intended for them, power consumption in idle mode is reduced.
{As MTDs avoid frequent wake-ups from listening to paging information not intended for them, the power consumption in idle mode reduces.}
{Thus}, a low-complexity wake-up receiver is an essential requirement~\cite{xia2019emerging}.~{{A paradigm shift from the traditional duty cycling (DC) medium access control (MAC) operation to on-demand WuR operation {is} envisaged {due to} the latter's energy efficiency superiority~\cite{froytlog2019ultra}.} It has been shown in~\cite{oller2015has} that a {WuR’s} average power consumption is around 1000 times lower than that of the main radio. Furthermore, the implemented WuR in~\cite{magno2016design} achieves around 70 times longer lifetime than DC protocols (with 1\% duty cycling) under light traffic load. With such potential energy savings, {WuR appears as a promising technique for achieving a lifespan beyond 10 years, which is the targeted lifetime for NB-IoT and 5G IoT devices.} Moreover, WuR enables instantaneous response to on-demand IoT data transmissions, resulting in much shorter latency~\cite{froytlog2019ultra}.}

Improving the efficiency of the standardized {DRX} mechanism for MTC has been the focus of much attention{, e.g.,} \cite{R7C2_1, ramazanali2016tuning, R7C2_2, R7C2_3, Adrx}. Authors in~\cite{R7C2_1, ramazanali2016tuning} discuss the tuning of a given set of DRX parameters in LTE/LTE-A and analyze the impact on energy-efficiency, while an improvement to DRX based on the inactivity timer management is provided in~\cite{R7C2_2}. Authors in~\cite{R7C2_3} propose {a} DRX mechanism {that exploits} the radio resource control connection release and re-establishment to save significant energy in MTDs, while an artificial intelligence (AI)-based adaptive DRX is devised in~\cite{Adrx}. {However,} since the DRX mechanism is not specifically designed for MTDs, more suitable {mechanisms are required}~\cite{canweachieve}.

%\textcolor{blue}{Note that demodulating a WuS is less demanding than standard NB-IoT network/communication signaling.}
Meanwhile, MTDs operating with WuR are considered in~\cite{R7C2_4,R7C2_5, mazloum2020interference, rostami2018wireless, rostami2019wake}. {Specifically,
authors in~\cite{R7C2_4} consider low-complexity WuS to improve energy-efficiency while {MTC re-synchronization} is carried out. The 
benefits of using passive WuR in wireless energy harvesting networks are highlighted in~\cite{R7C2_5},} whereas efforts to improve WuS in interference-free %orthogonal frequency division multiplexing 
OFDM based systems are made in~\cite{mazloum2020interference}{.   
%Note that since} demodulating a WuS is less demanding than standard %NB-IoT network/communication signaling, WuR should be optimized to significantly reduce the average energy consumption of an IoT device without increasing communication latency~\cite{odelberg20212}. %Nevertheless,
In all cases, WuS mechanisms use static operational parameters, which in \cite{rostami2018wireless, rostami2019wake} are} determined by the base station (BS) at the beginning of the session. %Meanwhile, artificial intelligence (AI)-based dynamic mechanisms for efficiently tuning WuS parameters in low-power MTC networks remain unexplored.}
{Remarkably, AI-based mechanisms for efficiently tuning WuS parameters in low-power MTC networks, although naturally appealing, remain unexplored. Therefore, we consider there is} vast potential for further reducing energy consumption by incorporating intelligence into WuS.

%\textchanged{Despite all the enormous efforts made by science in terms of energy-efficiency, as well as the advances achieved through techniques such as DRX and WuS, given the expected massification of MTDs and the imminent need to guarantee sustainable development also in the area of communications, it is essential to perfect the energy saving mechanisms that allow more and more to adjust power consumption to the actual operation requirements.}

%The main contributions of this paper are outlined next. 
\begin{center}
    \begin{table}[t!]
    \caption{Brief summary on benchmarks and our proposal}
    \label{comparison}
    \centering
    \begin{tabular}{lp{6cm}}
    \hline
        \textbf{Ref.}    &   {\textbf{Features}}\\
        \hline \\
        DRX~\cite{ramazanali2016tuning} &	Standard DRX mechanism with optimized parameters, Poisson traffic model, traffic analysis, target mean delay\\
        \\
        WuS~\cite{rostami2019wake}		&	{Standard WuS, predefined time interval between {low-power states with beacon seeker (WRx-On)}, Poisson traffic model, traffic analysis, target mean delay}\\
        \\
        FWuS      		                &	Low-power MTC traffic model using PPPs, AI-based traffic analysis, dynamic forecasting model to optimize WuS parameters, adjustable time interval between WRx-On states, target mean delay\\
    \hline
\end{tabular}
\label{comparison}
\end{table}
\end{center}

\vspace{-25pt}
{Herein, we aim at designing an efficient AI-enabled WuS mechanism. The main contributions of this paper are outlined next. First, we model the position of MTDs and event epicenters as distinct and independent Poisson point processes (PPPs), %as well as 
and 
%we 
define a function to model the influence of events on MTC traffic. {It is noteworthy that the model is able to characterize the different traffic patterns described in the literature{, e.g.,}~\cite{eldeeb2021learning, 6629847, lopez2021csi}%, since a periodic update pattern is much easy to predict
.} 
%We analyze the MTC traffic patterns using a simple neural network (NN) to build a predictive model named forecasting WuS (FWuS). 
We leverage a simple neural network (NN) to predict MTC traffic patterns and configure WuS accordingly. We refer to such mechanism as forecasting WuS (FWuS). }
Then, we use %the 
FWuS %model 
to optimize the wake-up parameters of the {MTDs}. In addition, the mean delay and power consumption {performance are} quantified for a set of wake-up parameters. Finally, numerical results evince the {superiority} of our proposed method with respect to {WuR~\cite{rostami2019wake} and DRX\cite{ramazanali2016tuning} based reference mechanisms.} 
%, whose features are summarized in Table~\ref{comparison}. 
{Table~\ref{comparison} summarizes the main features of the proposed solution and the benchmarks.} {To the best of our knowledge, optimizing WuS in this type of scenario has not been proposed before. Our work takes on significant relevance when considering the ever increasing number of connected MTDs, the urge for reducing
their energy consumption down to actual operation requirements, {and societal and economic goals regarding sustainable.}}

The rest of this paper is organized as follows. Section~\ref{sec3} describes the system model, while Section~\ref{newsec} overviews WuS technology and describes the traffic model. Section~\ref{sec4} formulates the problem and %proposes an energy efficient solving scheme
introduces the proposed scheme. Section~\ref{sec5} describes the framework for evaluating the system performance.
%under the proposed method. 
Numerical results are illustrated in Section~\ref{sec6}, and conclusions are drawn in Sections~\ref{sec7}. {Tables~\ref{Acronym} and~\ref{Symbol} list respectively %the 
acronyms and symbols used throughout this paper. %, 
{%however, 
For simplicity, we have omitted the well-known %terms
acronyms.}}

{\begin{center}
    \begin{table}[t!]
    \caption{Table of Acronyms}
    \label{Acronym}
    \centering
    \begin{tabular}{ll}
    \hline
        %\processtable{Table of Acronyms. \label{Acronym}}
        %{\begin{tabular}{@{}ll@{}}
        %\toprule
        {\textbf{Symbol}} &   {\textbf{Description}}\\
        %\midrule        
        \hline 
                {\textit{A}}			&	{Active state}\\
                %AI 				&	Artificial intelligence\\
                %BS      		&	Base station\\
                {DRX}     		&	{Discontinuous reception}\\
                {FWuS}		    &	{Forecasting WuS}\\
                {\textit{I}}              &   {Idle state}\\
                %IoT     		&   Internet of things\\
                {LSTM}			&	{Long short-term memory}\\
                %LTE-M		    &   Long term evolution for MTC\\
                %MAC 			&	Medium access control\\
                %mMTC		    &	Massive MTC\\
                {MTC}			    &	{Machine-type communication}\\
                {MTD}			    &	{Machine-type device}\\
                {NB-IoT}		    &	{Narrow-band internet of things}\\
                %NN 			    &	Neural network\\
                {PDWCH}		    &	{Physical downlink wake-up channel}\\
                {PO }			    &	{Page occasion}\\
                {PPP}			    &	{Poisson point process}\\
                %R				&	Regular state\\       
                {RFI	}			&	{Radio frequency interface}\\
                {RMSE }   		&	{Root mean square error}\\	
                {RNN	}		    &	{Recurrent neural network}\\
                %std			    &	Standard deviation\\
                {TTI	}		    &	{Transmission time interval}\\
                {WuR	}		    &	{Wake-up radio}\\
                {WRx-On}		    &	{Low-power state with beacon seeker}\\
                {WuS }			&	{Wake-up signal}\\
                %3GPP			&	3rd generation partnership project\\
                %5G      		&	Fifth generation networks\\
                %6G      		&	Sixth generation networks\\	
        %\botrule
        %\end{tabular}} {}
    %\end{table}
    \hline
\end{tabular}
\label{Acronym}
\end{table}
\end{center}
%\vspace{-10ex}
%\begin{center}
    %\begin{table}[t!]
    %\centering
    %\processtable{List of Symbols. \label{Symbol}}    
        %{\begin{tabular}{@{\extracolsep{\fill}}ll}\toprule
\begin{center}
    \begin{table}[t!]
    \caption{List of Symbols}
    \label{Symbol}
    \centering
    \begin{tabular}{lp{6.8cm}}
    \hline            
            \textbf{Symbol} &  \textbf{Description} \\ %\midrule
            \hline 
%			$c$             &       Coverage area of an MTD coordinator\\ 	
%			$d$           &         {Distance between an event and a sensor}\\
			$d$           &         {Distance between the place of occurrence of an event and the position of a sensor}\\
            $D$             &           Packet delay\\
            $\overline{D}$       	&           Mean packet delay \\
            $p(d)$						&			Distance function\\
            $g_i$						&			Predictions of inter-arrival time (expected values)\\
            $g_{o_{i}}$				&			Observed inter-arrival time (known results)\\
            $P_{A}$				&	Probability that a device goes to active mode\\
            $p_{f}$					&			False detection probability\\ 
            $p_{i}$						&			Probability of being in $S_i$\\ 
			$P_{ij}$			&	Transition probability from a state $S_i$ to $S_j$, $P(S_j \vert S_i)$\\
        	$p_\text{md}$		&			Miss-detection probability\\       
			$\overline{\text{PW}}$		&			Mean power consumption\\
			$\text{PW}_{i}$				&			Power consumption in $S_{i}$\\
			$\text{PW}_{b}$				&		Power consumption of a benchmark scheme\\
			%$PW_{2}$					&			Power consumption in $S_{2}$\\
			%$PW_{3}$					&			Power consumption in $S_{3}$\\
			%$PW_{4}$					&			Power consumption in $S_{4}$\\	
        	$q$				    	&			Geometrical distribution parameter\\
        	$R(k,s)$              &           Traffic rate in a time slot $k$ and state $s \in {I,A}$\\
			$S_i$						&			WuS state; $i = (1, 2, 3, 4)$\\
			$t_\text{mac}$					&			Delay due to the MAC protocol\\	
			$t_\text{on}$	&		On time\\%Time in ON state for cycle in TTI units\\
			$t_\text{pd}$					&			Power down time\\
			$t_\text{sleep}$		    		&			Forecasted sleep time\\
			$t_\text{u}$						&			Start up time\\
			$t_{1}$						&			WRx-On time\\
			$t_{2}$						& 			Active-decoding time\\
			$t_{3}$						&			Inactivity time\\
			$t_{4}$						&			Sleep time\\
%			$U_V$	                &	Variations experienced in the data that are not predicted\\			
%			$T_V$						&			Total variations in the data\\	
			$\eta$						&			Relative power saving\\
			$\eta_\text{max}$			&			Maximum relative power saving\\
			$\eta_\text{min}$				&			Minimum relative power saving\\	
			$\lambda_{E}$			&			Event density\\
			$\lambda_{M}$			&			MTDs' density\\
			$\Phi_{E}$		&	{PPP of event} epicenters\\	
			$\Phi_{M}$				&		{PPP of} MTDs deployment\\
        %\botrule
        %\end{tabular}}{}
    %\end{table}
%\end{center}
\hline
\end{tabular}
\label{Symbol}
\end{table}
\end{center}
\vspace{-10ex}}

%\section{Background}\label{sec2}
%
%\subsection{Wake-Up Scheme}
%
%\emph{\textbf{Profe Aqui no se si poner algo mas de estos temas o con lo que hablo despues es suficiente.}}
%\subsection{Long-Short term Memory}

\section{System Model}\label{sec3}
\begin{figure}[t!]
	\centering
	\includegraphics[width=\columnwidth]{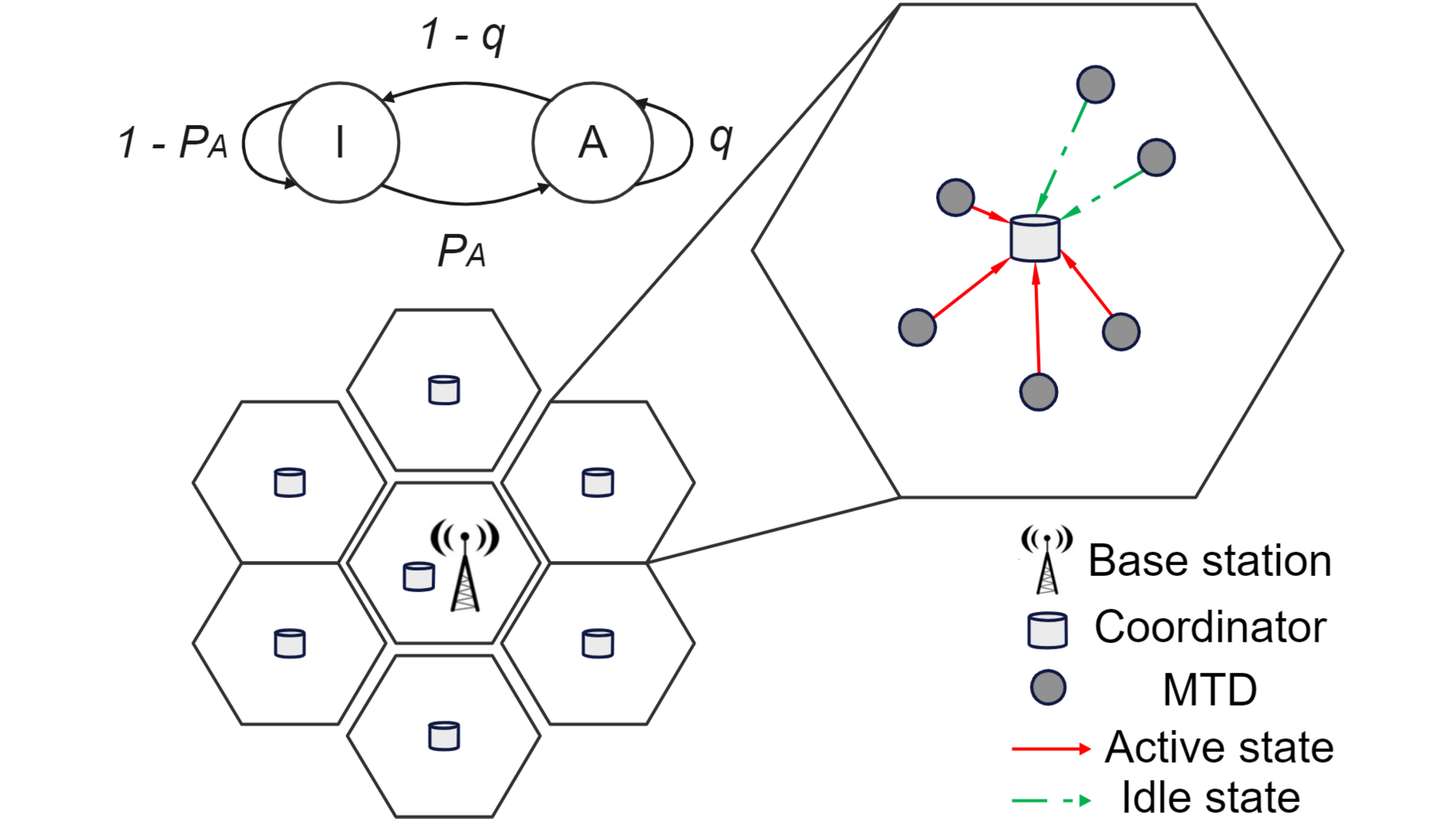}
	\caption{Illustration of an MTC network where the coverage area of the BS is split in several smaller  areas\protect\footnotemark[1]~in which a coordinator controls and collects information from the MTDs. Traffic exchanged between MTDs and the coordinator is modeled as a two-state complete Markov chain.}
\label{figure1}
\end{figure}

%
%This paper is framed in the system model  presented in Fig.~\ref{figure1}. 
We consider the coverage area of a single BS, where multiple coordinators{~\cite{8891507}} are deployed to serve as gateways of short-range MTDs, as depicted in Fig.~\ref{figure1}. This allows saving energy at the MTDs, which otherwise would need more power to communicate with the BS{, and serves as a solution to those MTDs that cannot directly contact the BS due to the use of short {range communication technologies~\cite{abdrabou2021application}.}} 
%Both 
The MTDs %and the coordinator 
are {equipped with a WuR} module to save energy {by allowing the radio frequency interface (RFI) to monitor the medium at regular intervals looking for triggering signals}. %However, we focus on the coordinator/MTD link since we assume that the coordinator has enough memory to store the information it obtains from the MTDs, while the behavior between a coordinator and the MTDs is similar to that of the BS with the coordinators of each cell.}
Each coordinator controls all the information exchange within its cell, {deciding whether to send a WuS that allows the MTD to %turn ON the radio frequency interface (RFI). 
exchange information.}

%Although several approaches\textcolor{blue}{, e.g.,}~\cite{shehab2020traffic, eldeeb2021learning, madueno2015reliable}, focus on the uplink channel, where

{Note that in some applications
the coordinator is in charge of 
%grant uplink schedule opportunities for 
granting/scheduling access to
MTDs that have requested it \cite{shehab2020traffic, eldeeb2021learning, madueno2015reliable}.
However, in some others, the MTDs need} to be attentive to the coordinator~\cite{rostami2019optimized, rostami2019wake}, {which may inform the MTD of} event anticipation or about an event that was not sensor-detected. In this way, MTDs need to be checking the physical downlink wake-up channel (PDWCH) regularly because, even when their sensors have not been activated by an event, the coordinator still 
%would need to 
may request information from them, or %anticipate 
for anticipating future activation for smooth applications like target tracking{.
%Following the previous cases and 
%Given the great advances in AI, a good approach would be a system where the coordinator controls the cell and would be able to activate specific MTDs when the situation needs it, even if they were not asking for a grant previously. With this scheme we would be able to use stronger forecasting systems in the coordinator that for the MTDs would be just impractical, due to the need to simplify their computational cost and restrict the consumption to prolong their battery lifetime. %An approach from the coordinator leaves all the hard work to the coordinator.
An} appealing approach may be that in which a coordinator activates specific MTDs based on AI-enabled forecasting mechanisms% even if the MTDs did not requested it
. Such mechanisms can be deployed at the coordinator side to keep the MTDs simple and energy-efficient. 
{Moreover, the coordinators can control the medium access, avoid collisions, and consequently reduce the energy consumed by the MTDs when they are not scheduled and would be performing idle listening~\cite{sachs20185g},~\cite{mahlouji2018analysis}.}  

%Observe that continuous
%searching for events in their covering area that trigger an alarm state. Continuous %monitoring of the covering area 
%traffic exchange between the MTDs and the coordinator
%causes an enormous energy consumption in both the MTDs and the coordinator.
%, wasting battery due the unnecessary amount of time in with the RFI is ON %active mode 
%at these devices, in the same way for link between the coordinators and the BS.
%WuS comes as a solution 

%to save energy~\cite{xia2019emerging} by allowing the radio frequency interface (RFI) to monitor the medium at regular intervals looking for triggering signals indicating the beginning of an information exchange period. %monitoring the medium at regular intervals in search of triggering events.
{For the sake of contextualization, consider} the following potential scenarios: {i)} target tracking, and {ii)} visitors of a historic arena. In the first case, MTDs that detect an event may trigger the surrounding nodes, and a set of MTDs activated by the coordinator may track/localize the target. Activating only the sensors close to the path followed by the target is a reasonable approach. The awakened group of MTDs may localize the target accurately. %Consider another use case: visitors of a historic arena. T
In the second scenario, the objective of the MTC network is to detect the visitors who are trying to enter forbidden areas. During the open hours of the museum, the rate of these events may be larger than during the night shifts. Daytime operation of the network can be based on periodic WuS scheduling, {whereas nighttime operation can follow a more suitable wake-up approach.}
\drg{\footnotetext[1]{{This paper focuses only on intra-cell communication. Note that cells do not have to be precisely hexagonal or the same size, since the MTDs  deployment density
(MTDs/$m^2$) depends on the area and not on the shape of the region of interest. We adopted the current representation for aesthetic reasons and in analogy to the classical literature on mobile communications.}}}
\subsection{MTDs Deployment and Event Sensing}

%We denote as $c$ 
Consider a {single} coordinator and {let 
%$N$ 
the} MTDs be deployed randomly and independently in its coverage area. %with radius r 
%n $c$
%, notice that
%we deploy $c$ as hexagon form just for illustration purpose,
%the proposed method is independent of the coverage area form since the MTD density deployment (MTD/$\mathrm{m}^2$) depends on the area value itself and not on the form. 
{We resort to homogeneous PPPs to model the position of devices and event epicenters \cite{thomsen2017traffic}. 
%typical nodes and events can be reasonably assumed to be stochastically deployed in the plane \cite{thomsen2017traffic}. %furthermore allowing for analytic tractability. 
%A function which models the influence of events on device traffic is also defined. 
Note that} PPP has been the most popular spatial model for various types of wireless networks~\cite{chen2020energy,deng2020energized} %as the PPP model has several convenient features, such 
because of salient features such
as the independence between points and the simple form of the probability generating functional~\cite{li2020resource}.
%
%The coordinators are deployed at fixed positions while the MTDs are deployed in the coverage area of their corresponding coordinator according to a homogeneous 
{The PPPs of the MTDs positions and event trigger epicenters in the Euclidean plane are denoted respectively by $\Phi_{M}$ and $\Phi_{E}$.  These processes are assumed to be independent and
to have density $\lambda_{M}$ and $\lambda_{E}$, respectively.}

\begin{figure}[t!]
	%\centering
	\centerline{\includegraphics[width=1.1\columnwidth, height = 4.9cm]{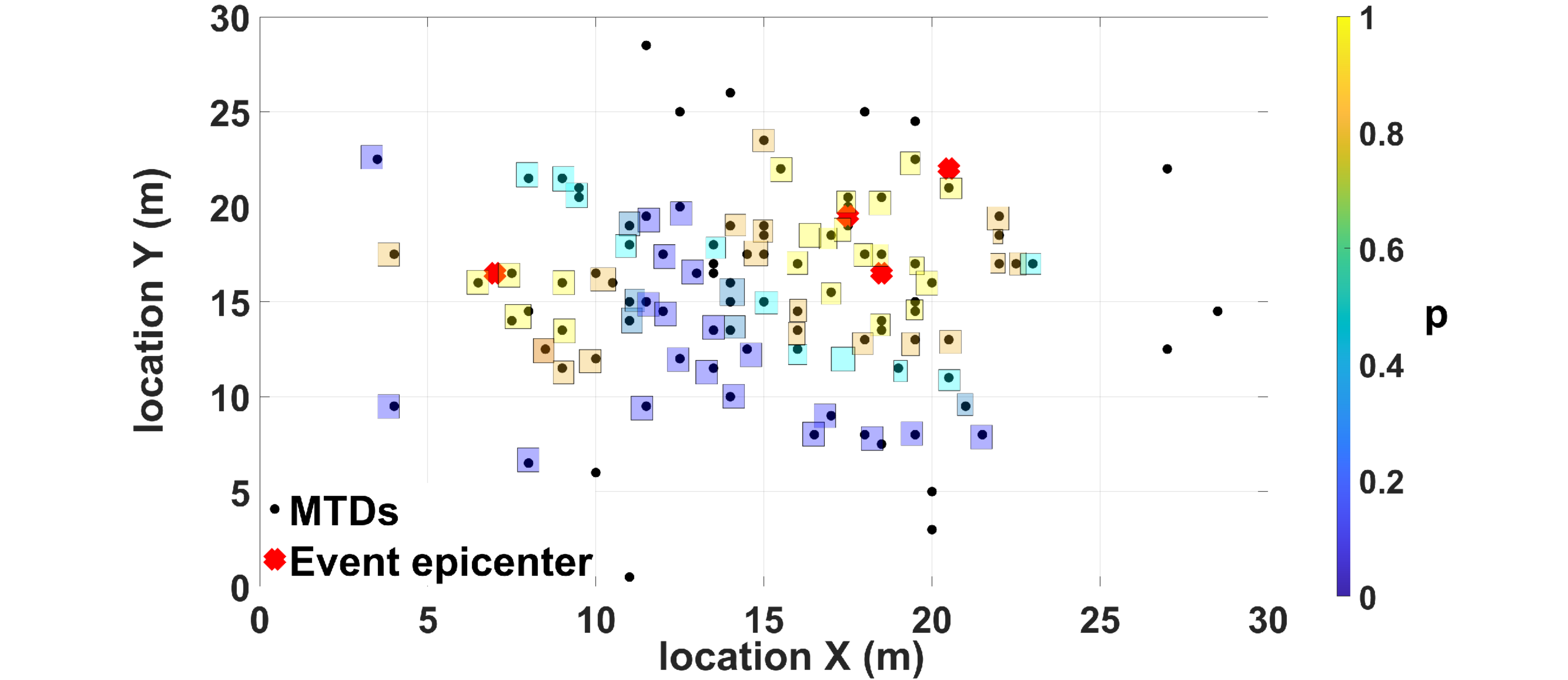}}
	\caption{Illustration of the influence of events on the surrounding MTDs{. 
	%, modeled as a probability function that depends on the distance between the MTDs and the event epicenter. 
	Colored} squares are related to the event activation probability in the colorbar (right). The absence of a square means zero activation probability.}
\label{figure3}
\end{figure}

%In order to capture the effect of a given event on an MTD, 
{Let $p(d)$ denote} the probability that {a certain MTD senses an}  event with epicenter at {a distance $d$} in the Euclidean plane $\Re^{2}$. 
%This function depends on the distance $d$ between the two locations, 
{Then,} we have $p(d): [0, \infty) \rightarrow [0,1],$
%\begin{equation}
%\label{eq1}
%	f: [0, \infty) \rightarrow [0,1],
%\end{equation}
where $p(d)$ is typically non-increasing to represent a decaying influence of events  as the distance $d$ increases. Fig.~\ref{figure3} %depicts the effect of events occurring as a function of distance between the MTD and the event epicenter.
depicts an instantaneous realization of the MTC network and event epicenters.
\subsection{MTD State Modeling}
Each MTD has sensors that trigger an interrupt if they detect an event (e.g., a peak in energy consumption or a relevant variation in energy consumption). {Upon detecting an event, the MTD connects the necessary parts to process this signal, which had been turned OFF to save energy~\cite{zhang2020opportunities}.} The information is stored in the MTD awaiting an RFI activation signal from the %cell 
coordinator. The MTD measures and processes, but does not transmit until indicated by the coordinator. %In this way, the MTD does not miss any event detection. % in case the WuS received from the coordinator does not arrive on time.%{\footnotemark[2].} %Likewise, each coordinator is subordinated in the same way to the BS, each coordinator receives and processes the information received by the MTDs, however it does not transmit to the BS until receiving a WuS from the BS itself. %The WuS and its characteristics will be modeled in Section~\ref{sec31} while in Section~\ref{sec32} the characteristics of the traffic obtained in our system will be modeled.

%{\footnotetext[2]{ {Likewise, each coordinator is subordinated in the same way to the BS, each coordinator receives and processes the information received by the MTDs. However, it does not transmit to the BS until receiving a WuS from the BS.}}}

%We take c as the coverage area (hexagon) of a machine type device (MTD) coordinator of radius r and let N MTDs be deployed randomly and independently in c. 
%Let a BS, which controls the traffic in a specific area, be divided into cells, and in each cell, several  (MTDs) be controlled by a centered MTD coordinator which in turn send information of those MTDs to the BS as shown in Fig.~\ref{figure1}. 

Each MTD can be in one of two states: 
\begin{enumerate}
    \item idle (\textit{I}): {the MTD wakes up (ON) the RFI at regular intervals, while it waits %a triggering event, either sensor-detected or 
    a WuS from the coordinator.}  %looking for a wake up signal, 
%the device wakes up (ON) at regular intervals monitoring the environment waiting for an event trigger 
%in which case a transition to active state is needed. 
The MTD may experience a sensor-detected interrupt, however, that information is stored and the MTD remains in state \textit{I} until receiving a WuS.
The MTD saves considerable energy as its {RFI is OFF} most of the time.

    \item active (\textit{A}): the device exchanges information with the coordinator.  %\rds{Mas em regular ele não comunica com o coordinator, comunica?}. 
    The RFI is {continuously} ON, which translates into high energy consumption.  
\end{enumerate}
%
%The transition from R (idle) to A state occurs when an event is detected \rds{Parece em conflito com o início desta seção, não é quando recebe o WuS? É uma junção entre ter detectado um evento e ter recebido um WuS. Ou ele muda para Active quando detecta um evento e transmite só quando recebe um WuS? Acho que precisa ficar bem claro o seguinte: O que o device faz e não faz em cada estado; como o device muda de um estado para o outro. Está um pouco confuso.}, while the MTD stays at A state for the duration of the event. 
The transition from state \textit{I} %(idle) 
to \textit{A} occurs when %after the detection of an event at the MTD 
a WuS is received from the coordinator, thus enabling the information exchange between {the MTD and the} coordinator.
Upon receiving the WuS, the MTD goes to state \textit{A}, either because it detected an event through its sensors and {the coordinator granted a} communication authorization (uplink grant) through the WuS, or because the {WuS informs} it of an imminent predicted event (request to transit to state \textit{A}). 
%enabling the information exchange between the two, the coordinator and the MTD, while the MTD stays at A state for the duration of the event.
%The coordinator sends a WuS enabling the information exchange between the MTD and the coordinator itself, while 
The MTD stays at state \textit{A} for the duration of the event.

%In regular state 

%and the MTD stays at this state for the duration of the event. 

%Fig.~\ref{figure1} shows an example of a deployment of the system model in a covering area by a BS. The system is composed by a BS coordinating several cells, each one with its respective MTD coordinator in order to compile the information of each cell and tribute to the BS. 
{The triggering {events 
%\textcolor{blue}{that activate the} MTDs 
%in covering area  
change the MTC} traffic characteristics towards the coordinator.} In Fig.~\ref{figure1}, four MTDs are in state \textit{A} (indicated with solid red arrows) and two are in state \textit{I} (indicated with dashed green arrows). The MTDs wait for the WuS from the coordinator that would trigger state \textit{A}. {Note that according to the application, 
there are three elementary MTC} traffic patterns~\cite{zanella2014internet, bayindir2016path, ratasuk2015recent}: periodic update, event-driven and payload exchange. {Nevertheless, real} world applications often  combine these traffic types. Hence, considering the three elementary classes above enables building models with an arbitrary degree of complexity and accuracy~\cite{eldeeb2021learning, 6629847, lopez2021csi}. 

Finally, we assume time is slotted in transmission time intervals (TTI)%of 1 ms
. In slot $k$ and state $s\in\{I,A\}$, $\text{MTD}s$  generate traffic with rate $R(k,s)$, depending on their current state.
%In the regular state, it generates traffic with rate $R(k,R)$, and in active state, the rate is $R(k,A)$.
%In this work, we assume $R_i(k,s) = R(k,s)$, $\forall{i}$.% = 1, \ldots, N$. 
%Notice that this coverage hexagon is just one of many areas that tribute to the primary BS, see Fig.~\ref{figure1}.

\section{Traffic and Wake-Up Modeling}\label{newsec}
The traffic exchanged between the coordinator and the MTDs is modeled using an ergodic Markov chain with two states~\cite{thomsen2017traffic}, \textit{I} and \textit{A}, as illustrated in {Fig.~\ref{figure1}}. 
%\textchange{The model in~\cite{thomsen2017traffic} considers the event-driven pattern with an event density parameter.} 
The payload exchange pattern is modeled through the $q$ parameter in the geometric distribution. {This parameter $q$ considers how bursty could be the traffic generated by an event.} A device goes to state \textit{A} with probability
\begin{equation}
	P_{A} = 1 - \exp\left({-2\pi\lambda_{E}\int^{\infty}_{0}{p(d)\partial d}}\right).
\label{eq3}	
\end{equation}
%
%\noindent given a PPP $\Phi_{E}$ with density $\lambda_{E}$, where $f(d)$ is typically non-increasing,
%In this paper we use a negative exponential function $\left(f(d)=e^{-d}\right)$%(see Table~\ref{table1}). 
Once in state \textit{A}, the {MTD} remains {there} for a number $k$ of %time intervals 
TTIs, geometrically distributed with parameter $q$, thus, with probability mass function%represented in {the form} %\rds{o sentido físico dessa distribuição do tempo no estado A?}
\begin{equation}
	{f_{K}} (k)= (1-q)q^{k}, \ k = 0,1,\dots.%\in [0,\infty).
\label{eq4}	
\end{equation}%Notice that time ($k$) is measured in a unit small enough (TTI) so that it can be later expressed in integer values.

The introduction of the additional parameter $q$ in the Markov chain model allows  tuning the temporal correlation of the individual rate processes of the MTDs and, as a result, that of the total rate process. Hence, the parameter $q$ of the Markov chain enables tuning the model to suit various MTC applications and event reporting strategies. Furthermore, the parameter $q$ can be used to study the impact of events on the temporal correlation of the total traffic at the coordinator. Focusing first on the case of small values of $q$, %\textit
{e.g.,} $q \leq 0.1$, {the traffic behaves similarly to a Bernoulli process (memoryless)~\cite{thomsen2017traffic}, since it has a short memory.} Increasing the value of $q$ increases the memory, %\textit
{i.e.}, the total rate at a given time $k$ is correlated with many past values. This is because %the Markov chain now stays in state \textit{A} longer when it enters that state
once state A of the Markov chain is entered, one stays there longer~\cite{thomsen2017traffic}.

%\begin{equation}
%	f_{x} (x)= q^{(x-1)}-q^x, \ x \in (0,1).
%\label{eq4}	
%\end{equation}

%\noindent where $p = (1 - q)$. 
%All devices are set to regular mode at the beginning. 
We emphasize that, although modeled as in Fig.~\ref{figure1}, the traffic that arrives to the coordinator has parameters that are different from the one seen by each MTD since the coordinator combines the traffic from several MTDs. In fact, the MTDs could have traffic patterns with parameters that are different from each other. {For simplicity, we consider that the MTDs use a medium access control (MAC) technique to avoid collisions at the coordinator side.} Since the proposed method focuses only on the coordinator, it is independent of the resolution of this problem at the MTD level and the additional complexity that this entails. %Furthermore, it is noteworthy that the traffic model used in this paper foresees these constraints. \rds{não entendi a última frase}
\subsection{Wake-Up Scheme}\label{sec31}
\begin{figure}[t!]
	\centering
	\includegraphics[width=\columnwidth]{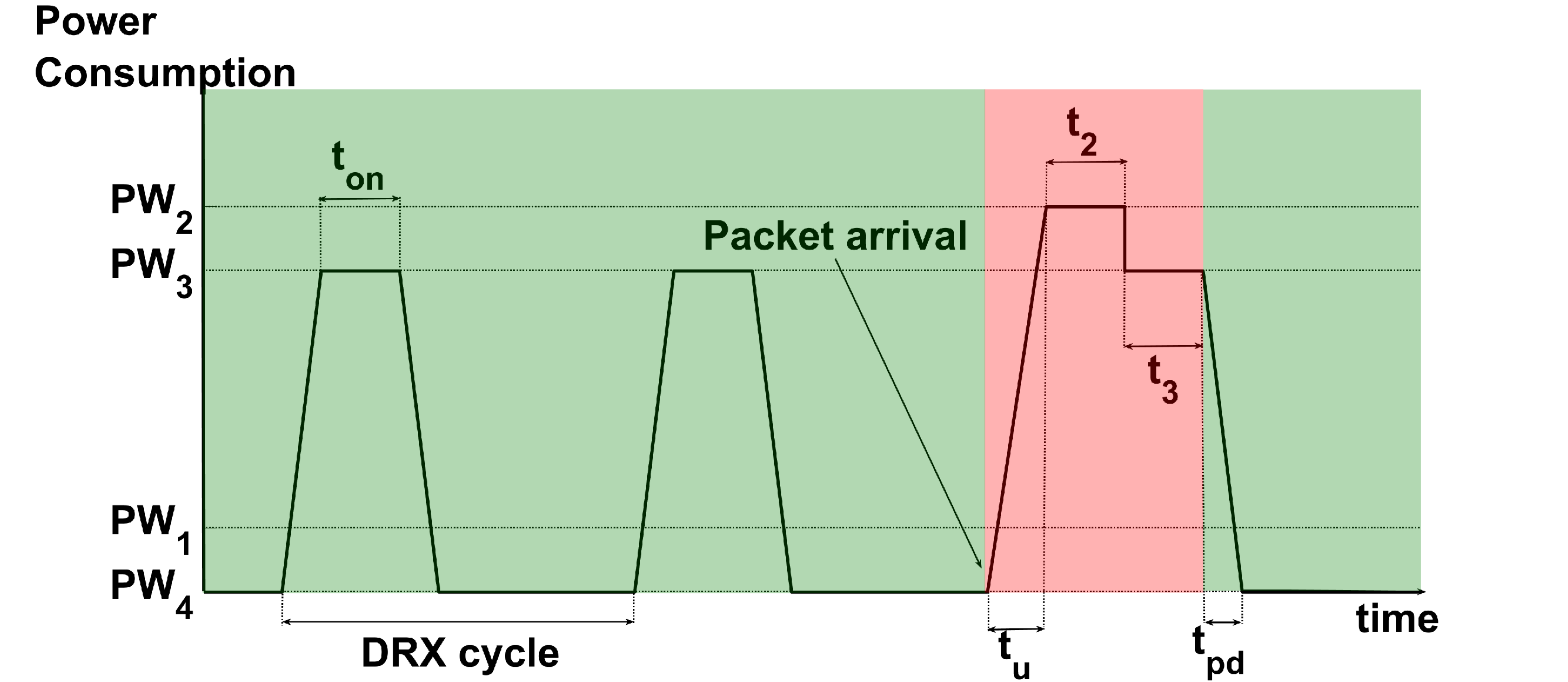}
    \includegraphics[width=\columnwidth]{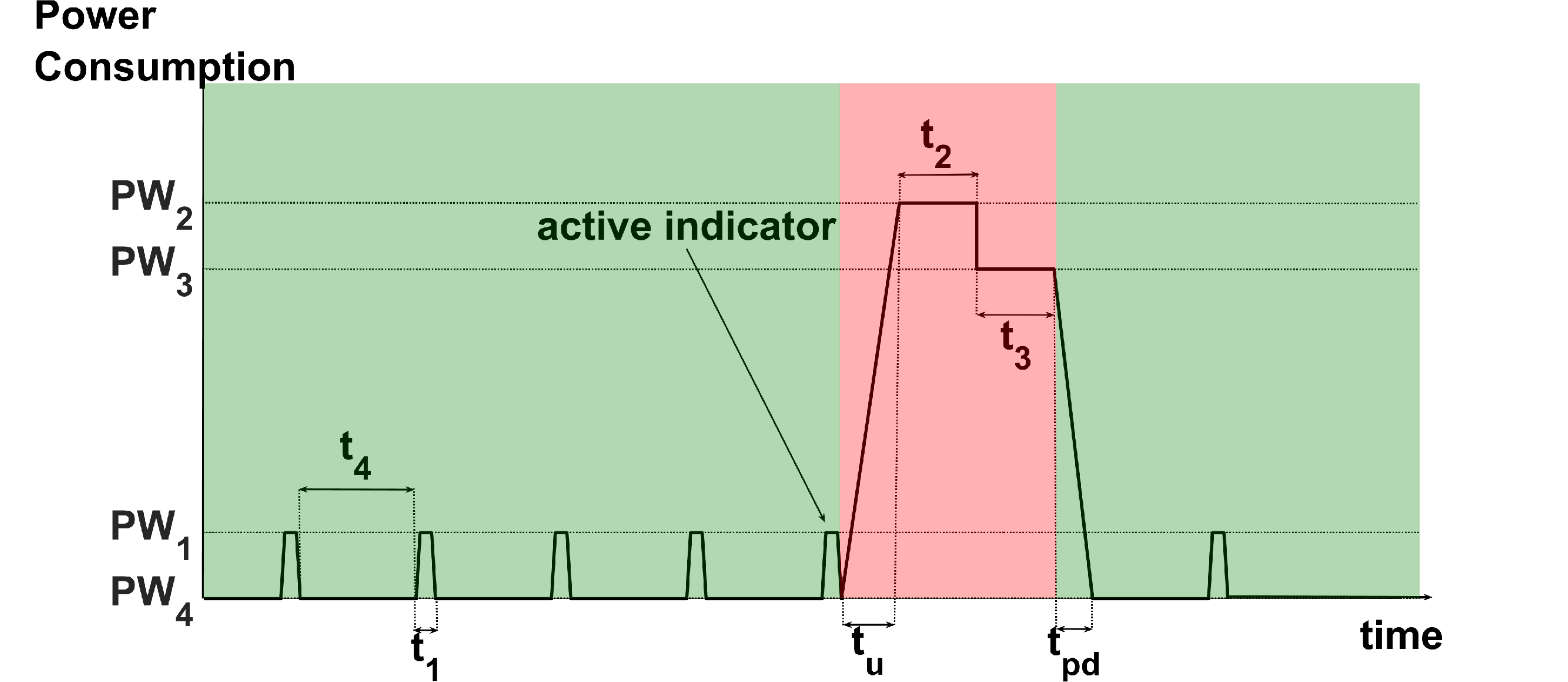}
    \includegraphics[width=\columnwidth]{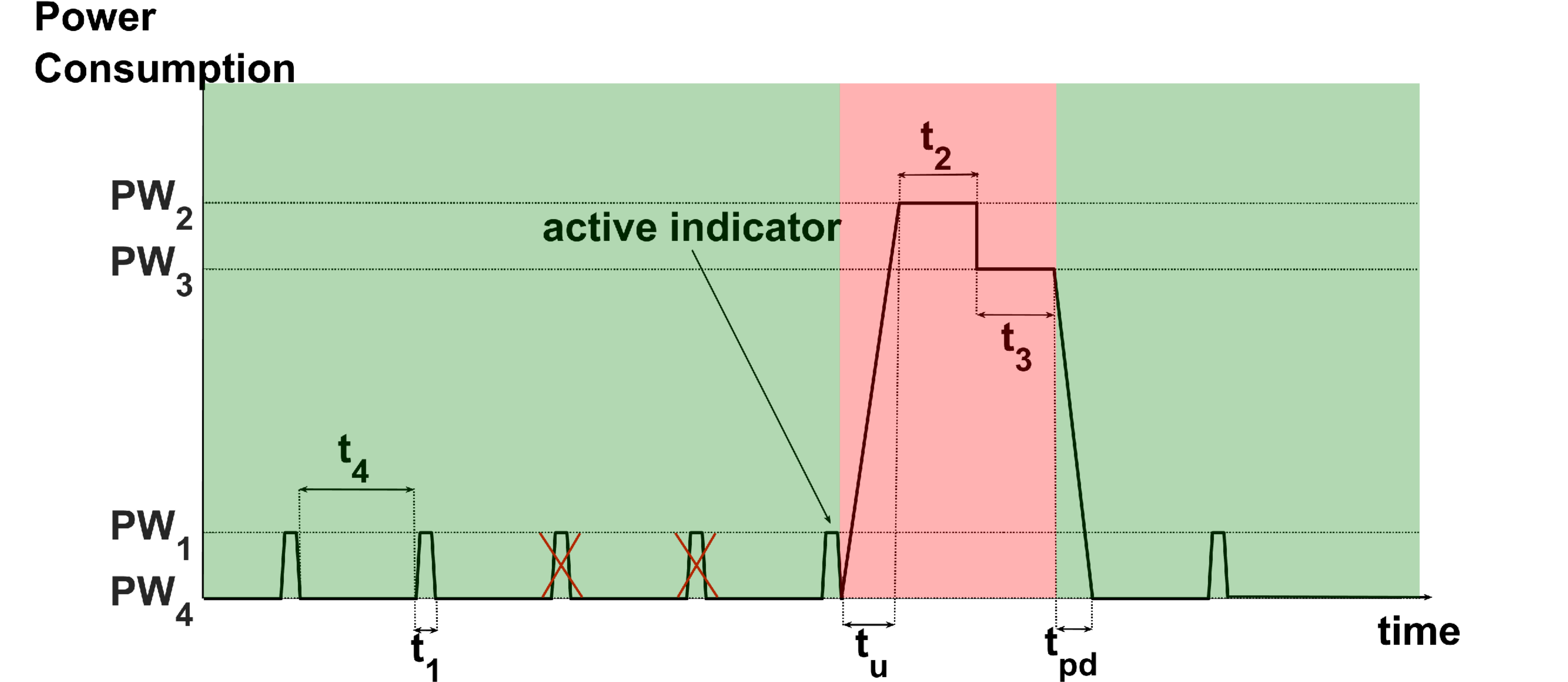}
    \includegraphics[width=0.7\columnwidth]{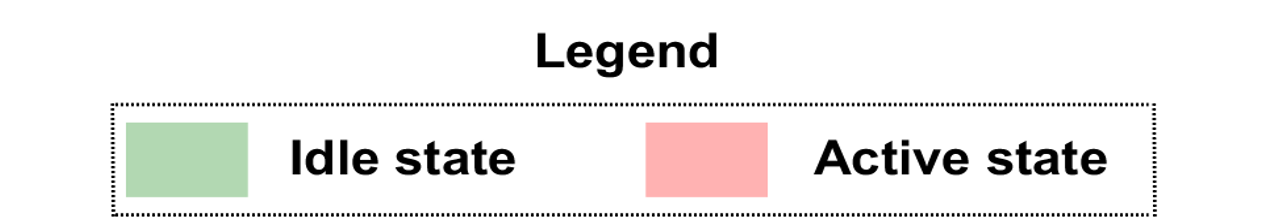}
    \caption{Illustration of the different consumption profiles for the proposed and the benchmark schemes. a) DRX mechanism consumption profile as described in~\cite{ramazanali2016tuning} {(top), b) different} wake up states and their respective power consumption (middle){, and c) proposed} FWuS, where turning ON the RFI {during WRx-On} events is avoided based on forecasting inter-arrival time (bottom){. In case of a),} PW1 is indicated only as a reference since this profile never works at this consumption level.}
\label{figure2.1}
\end{figure} 
{Idle listening at the MTDs is avoided by leveraging WuS technology. % by transmitting a compact signal in a configurable time before a device is paged.}
%WuS consists in a compact signal transmitted in a configurable time before the PO when a device is being paged, allowing the MTD to maximize its sleep time during periods when there is no paging. 
The coordinator sends a WuS to opportunistically} instruct the MTD that it must monitor for paging. Otherwise, the MTD skips the paging procedures, thus, it can potentially keep parts of its hardware switched off most of the time~\cite{dian2020lte}. 

{Fig.~\ref{figure2.1} depicts the different states and power consumption profiles for the proposed method as well as the benchmarks used for comparison purposes. Fig.~\ref{figure2.1} a) illustrates the DRX states as depicted in~\cite{ramazanali2016tuning}. Notice that  {the level $\text{PW}_1$ in DRX is not reached by any state. Conversely}, the consumption in ON state is comparable with that in the active-decoding state even when {no packet arrives.}
%packet arrival is not intended. 
%Now
However, by using WuS, the device {can} activate a low-power beacon seeker whose consumption ($\text{PW}_1$) saves energy with respect to the ON state in DRX, as shown in Fig.~\ref{figure2.1} b).} 
Note that the MTD is allowed to traverse to a {WRx-On state} waiting for an active indicator within the aforementioned {compact WuS} (beacon) which reveals the PO (${t}_2$). Such beacon communicates to the MTD the necessity of commuting to a {decoding state where} the consumption is higher. This switching occurs at regular intervals and has a duration ${t}_1$, while the device remains in sleep state between them to save energy. At the end of the {decoding} state, if no more packets are to be exchanged, the device waits for an inactivity time (${t}_3$) before returning to the sleep state. 
%\vspace{-5pt}

{Observe that despite the potential decrease in power consumption triggered by a WuS scheme, some idle states may  still occur where no WuS is detected.} In this paper, we aim to avoid these states (marked with red crosses) based on a forecasting model, which is presented later in {Section IV.} 
%The MTDs \textcolor{blue}{are 
%assumed to use WuS in the reception module to save energy. They are 
%equipped with a WuR} dedicated for channel listening. \textcolor{blue}{WuR is a potential} solution to reduce the idle listening energy cost and achieve lower delays, as demodulating the WuS signaling is less demanding than standard NB-IoT signaling~\cite{odelberg20212}.
%
\subsection{Traffic Model}\label{sec32}
\begin{figure}[t!]
	\centering
	\centerline{\includegraphics[width=1.1\columnwidth]{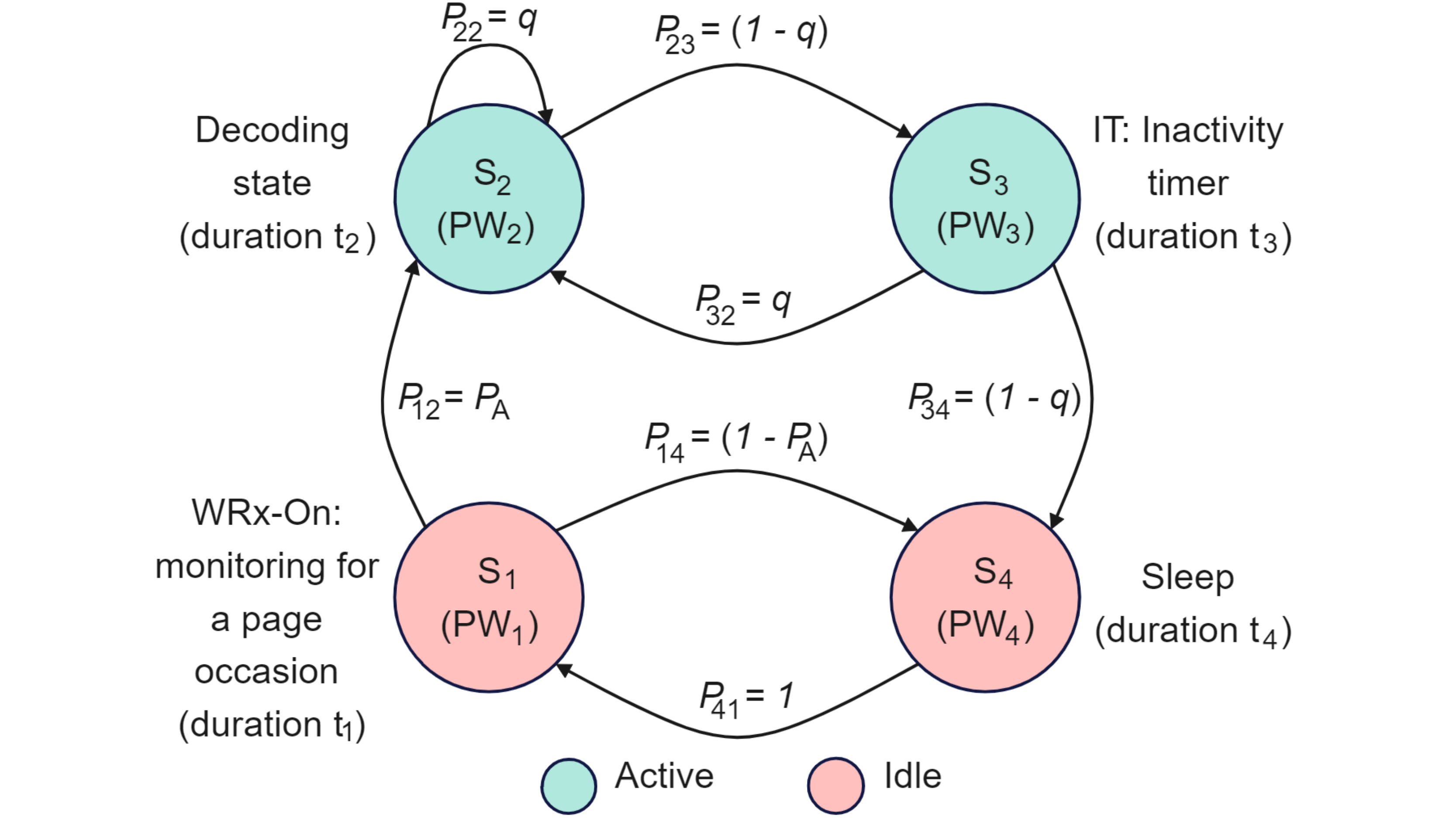}}
	\caption{{The wake-up} model as a semi-Markov chain. %\textcolor{red}{place $S_i$ in the figure states also.}
	}
\label{figure2.2}
\end{figure}

The transitions between {wake-up} states are modeled with a 4-state Markov chain as illustrated in Fig.~\ref{figure2.2}.
$S_{1}$ represents the WRx-On, at which the MTD monitors PDWCH looking for a WuS alerting of a PO. $S_{2}$ is the active-decoding state where the MTD decodes the information packets, while $S_{3}$ represents the state for the inactivity timer. Finally, $S_{4}$ corresponds to the sleep state in which the device is not able to receive any signal. 

The MTD remains in state \textit{I} for a time $t_{4}$, {while switching to $S_1$ inmediately after}. At $S_{1}$, the MTD monitors the channel searching for a WuS for a fixed time $t_1$. If the MTD detects a PO, it goes to $S_{2}$ {to decode the received information}. {If the whole} information arrives before $t_{2}$ ends, the MTD goes to $S_3$, otherwise $t_{2}$ is reset and the MTD remains in $S_2$. At $S_{3}$, the device waits for a PO for a time $t_{3}$ (inactivity timer). If a scheduled PO is received within $t_{3}$, the MTD returns to $S_{2}$, otherwise it goes to $S_{4}$ completing the wake-up cycle. Notice that the device consumes a start up time ($t_\text{u}$) in transitions from $S_1$ to $S_2$, and a power down time ($t_\text{pd}$) from $S_3$ to $S_4$. {The transition probability from state $S_{i}$ to $S_{j}$ is denoted as 
	$P_{ij} = P(S_{j} \vert S_{i})$
%\label{eq2}	
%\end{equation}
%
and also given in Fig.~\ref{figure2.2}.}
%, which are considered in our mathematical analysis. 
%Values for $t_1$, $t_2$, $t_3$ y $t_4$ are not fixed, 

In this paper, we aim to optimally configure $t_1$, $t_2$, $t_3$ and $t_4$ as to minimize the energy consumption {of the MTDs}. % 
\section{Proposed FWuS Scheme} \label{sec4}

The traffic of most MTDs is substantially different from human traffic accessing the Internet\cite{Adrx}. In most IoT use cases, the traffic is either generated periodically,  or as a burst after the detection of events \cite{zhou2019online}.
%It would be very challenging for the network to manage the burst traffic generated by many IoT devices in reaction to a specific event~\cite{dian2020lte}. 
%Compared with that of human type communication (HTC), the traffic patterns of MTC could be very bursty and even nonstationary~\cite{zhou2019online}.
%
Nonetheless, MTC traffic can be correlated in time and space\cite{graph}. This is because of the correlated temporal and spatial characteristics of a variety of events, e.g., low water or mineral concentration in crop soil and smart electricity metering. This sort of behavior can result in traffic bursts which can put stress on the receiver, e.g., the coordinator,  since MTC traffic is usually uplink-dominated. It is therefore desirable to have tractable models that can be used to assess the impact of this type of traffic at the receiver~\cite{thomsen2017traffic}.

In this context, accurate traffic prediction {mechanisms are appealing, specially those relying on AI as they can potentially capture the inherent dynamics and non-linearities of the system \cite{graph, hybrid_deep, dmTP}.} 
%be critical for efficient green networks~\cite{graph, canweachieve, hybrid_deep}. 
%In fact, AI-based traffic prediction is an emerging key block\cite{graph, hybrid_deep, dmTP}.
%, however, state-of-the-art prediction model results still have significant gap to the theoretical upper bound, which motivates novel prediction models and algorithms %capable of make full use of the hidden information from past data
%\cite{canweachieve}.
{Interestingly, 
%recurrent} NN (RNN) is an appealing AI technique for predicting the upcoming value of a given sequence~\cite{wang2017spatiotemporal}. \textcolor{blue}{Meanwhile,} 
long short-term memory (LSTM) is a popular type of recurrent NN (RNN) \cite{wang2017spatiotemporal}} that is specially designed to learn long-term dependencies of a sequence,
%for predicting the upcoming value of a sequence. 
%The term long-term dependency refers to the sequence, whose prediction results
thus, predictions are made based on long-sequences of previous input values rather than on a single previous input value~\cite{memon2019artificial, graph}. %\textcolor{red}{Therefore, short-term traffic flow forecasting model for large-scale traffic data in massive IoT is expected to make a high contribution to active traffic management~\cite{wang20206g, eldeeb2021learning}.}
{Different from classical solutions, e.g., relying on dynamic programming~\cite{bellman2015applied} and reinforcement learning~\cite{luong2019applications}, LSTM-based techniques~\cite{xu2021generative,CRAN,hybrid_deep} provide autonomous decision-making and relative fast} learning speed, especially in %the 
problems with large state and action spaces. 

%In this paper, we propose a scheme to %characterize the MTC traffic patterns, using an LSTM %RNN
%predict the MTC traffic pattern using an 
{We propose an LSTM-based traffic forecasting model and a methodology for properly configuring} the WuS parameters. 
%Fig.~\ref{figure4}} represents the diagram of an LSTM unit. Networks composed of LSTM units are RNN networks with extended-memory capabilities. LSTM cells allow the NN to remember its inputs over a long period of time. This memory is similar to that of a computer since the LSTM cell can read, write and erase information in its memory.
{Note that an} LSTM cell is made up of three gates, the input gate, the output gate and a forget gate. These gates determine if the information is read (input gate), if it is not relevant and is disregarded (forget gate), or if it is saved, impacting the current time step (output gate){.\footnote{{Interested readers in the details of LSTM networks are advised to review~\cite{gers2000learning, memon2019artificial}.}} %Interested readers in LSTM working operation are advised to review~\cite{gers2000learning, memon2019artificial}, as the LSTM network is a well-known recurrent NN architecture with a predefined heuristic and this analysis is out of the scope of this paper.
The} inputs are %the latest received packet and the feedback from the network are used
the time-stamp of the latest packet ($g_{o_i}$) and the latest relevant inter-arrival {time, which are obtained via feedback. 
The} inputs are modulated and used to train the unit. The forget {gate} contributes to dismiss over-training due to fluctuation in the training set. {Then, %we obtain 
a prediction for the next outcoming packet ($g_{i+1}$) is obtained as output.}
{As a result, we obtain a trained model {for} predicting the inter-arrival time between information packets %using the traffic data
using the known traffic data as input. Notice that the events that cause traffic data are independent among one another.}

{Recall that we aim to  configure the WuS parameters as to minimize the MTDs' energy consumption based on the modeled traffic rate and the given application. Now, observe that 
$t_3$} greatly impacts the energy consumption while the delay greatly depends on $t_4$. %Optimizing the values of $t_{4}$ and $t_{3}$ 
%set to a defined fixed delay value 
%is challenging, since it is not jointly convex. 
Tuning $t_3$ and $t_4$ to {minimize the MTDs'} energy consumption while maintaining a delay threshold is challenging since energy and delay are not jointly convex on $t_3$ and $t_4$.
%We optimize the values of $t_{4}$ and $t_{3}$ set to a defined fixed delay value for the packets, depending on the modeled traffic rate and the established use application. 
%These values are calculated for the proposed traffic model and 
{Herein, we resort to setting the value of $t_{3}$ to} 1 TTI to assure low power  consumption, {while} $t_{4}$ is optimized in a way that {allows satisfying} a mean delay constraint ($\overline{D}$):
\begin{equation}
	t_{4} = \max_{\substack{t_{3}=1 \ \text{TTI} \\ D = \overline{D}}} {\text{sleep}.}
\label{eq5}	
\end{equation}
{In \eqref{eq5},} $D $ is the packet delay and $\text{sleep}$ is the time between every channel monitoring state ($S_{4}$). 
\begin{figure}[t!]
	\centering	
	\centerline{
	    {%\tikzset{fontsize=10} 
	    \begin{tikzpicture}[node distance=2cm]

        \node (Data1) [startstop, xshift=2.5cm] {Latest Received Packet};
        %\node (in1) [io, right of=Data1, xshift=2.5cm] {Input};
        \node (pro1) [process, below of=Data1, font=\fontsize{9pt}{9pt}\selectfont, text centered] {Forecasting Model};
        %{\node (dec1) [decision, below of=pro1, yshift=-0.4cm, text width=2.5cm, font=\fontsize{10pt}{10pt}\selectfont, text centered] {\{$t_{\text{sleep}}/t_4$\} < 0.5?};}
        \node (pro4) [process, right of=pro1, xshift=2cm, text width=2.5cm] {
        %\begin{enumerate}
            %\item Set $t_3 = 1$ TTI;
            %\item Calculate $t_{4_{0}}$ using~\eqref{eq8}; 
        %\end{enumerate}}
        \textbf{1)} Set $t_3 = 1$ TTI        
        
        \vspace{1mm}
        
        \textbf{2)} Calculate $t_{4}$ \\
        \hspace{0.3cm} using~\eqref{eq5}% and~\eqref{eq8}
        };
        %{\node (pro2) [process,below of=dec1, yshift=-0.4cm, font=\fontsize{10pt}{10pt}\selectfont, text centered, text width=2.8cm] {$n$ = floor($t_{\text{sleep}}/t_4$)};}
        %\node (pro3) [process, right of=dec1,xshift=2cm, text width=2.6cm] {%\textbf{1)} %Calculate $n$ \\ ($n\times t_{4} \approx t_{\text{sleep}}$ \\  $\forall{n}=1, 2, \ldots$)
        %If \{$t_{\text{sleep}}/t_4$\} < 0.5, then\\
        %{\hspace{0.5cm}}$n$ = floor($t_{\text{sleep}}/t_4$),\\
        %else \\
        %{\hspace{0.5cm}}
        %{$n$ = ceil($t_{\text{sleep}}/t_4$)}

        %\vspace{1mm}
        
        %\textbf{2)} $t_{\text{sleep}} = n\times t_{4}$
        %};
        %$n\times t_{4_{0}} \\ (\forall{n}=1, 2, \ldots)$};
        
        \node (out1) [io, below of=pro1, yshift=-0.3cm, text width=2.7cm, font=\fontsize{9pt}{9pt}\selectfont] {{{$t_{\text{sleep}} =$ \\
        \vspace{1mm}
        $round(t_{\text{sleep}}/t_4)\times t_{4}$}}};
    
        \draw [arrow] (Data1) -- (pro1);
        \draw [arrow] (pro4) -- (pro1);
        %\draw [arrow] (in1) -- (pro1);
        \draw [arrow] (pro1) -- node[anchor=west] {$t_{\text{sleep}}$}(out1);
        %\draw [arrow] (dec1) -- node[anchor=west] {yes} (pro2);
        %\draw [arrow] (dec1) -- node[anchor=south] {no} (pro3);
        %\draw [arrow] (pro2) -- node[anchor=south] {$n$}(out1);%node[anchor=west] {$t_{\text{sleep}}$}(pro3);
        %\draw [arrow] (pro3) -- node[anchor=west] {$n$}(out1);
        %\draw [arrow] (out1) -- (Data1);
        %\node [below=1cm, align=flush center,text width=8cm] at (pro2){(b) Flowchart.};
    
    \end{tikzpicture}}}
    %\centerline{\includegraphics[width=0.98\columnwidth]{img/flowchart2.pdf}}
	\caption{Flowchart of the proposed WuS optimization{, $round(\cdot)$ represents the rounding operation to the nearest integer.}}
\label{flowchart}
\end{figure}
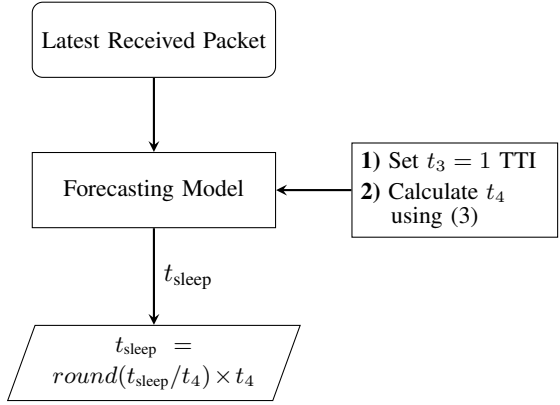
Depending on the prediction, the system is able to decide whether to enter state $S_{1}$ from $S_{4}$ or not. Fig.~\ref{figure2.1} c) illustrates the proposed FWuS scheme, while Fig.~\ref{flowchart} {shows} the proposed optimization flowchart. 
Once the WuS parameters are defined, they cannot be changed due to synchronization issues. However, we can extend $S_4$ up to $n\times t_{4}\ (\forall{n}=1, 2, \ldots)$ as in {Fig.~\ref{figure2.1} c).} %where the second $S_4$ has a duration of $3\times t_{4}$
%, (notice that I put $t_{4}$ as simplicity only to show the situation, because in reality this duration is $t_{sleep} = (n-1) t_{4} + (t_{4} - t_{1}))$.}
The system can dispense with WRx-On events, depending on the prediction, since it considers that the probability of %an imminent 
a packet arriving is very low. 
%note that the duration $t_{4}$ is predefined between the MTD and the coordinator. 
Since the MTDs have limited processing power, the coordinator %is in charge of running 
runs the forecasting {model and configures the WuS parameters.}
{%Using the equations (\ref{eq5}) and (\ref{eq8}) we calculate $t_4$, while the forecasting model gives us the proposed $t_{\text{sleep}}$ according to the data. Now, a quick comparison is made, if the time proposed is below $+1.5\times t_4$ then, the time between $T_{\text{ON}}$ of duration $t_1$ is equal to $t_4$. On the other hand, if the $t_{\text{sleep}}$ proposed by the forecasting model is above $+1.5\times t_4$ then, we calculate how many steps of duration $t_4$ fit in the $t_{\text{sleep}}$ duration, therefore, the time between $T_{\text{ON}}$ of duration $t_1$ is equal to $n\times t_4$, being $n$ the number of times $t_4$ fits in $t_{\text{sleep}}$. Notice that, even when the proposed $t_{\text{sleep}}$ is not equal to $t_4$ the WuS parameters were prearranged between the coordinator and the MTDs, therefore in order to maintain the synchronism between the coordinator and each MTD the time between $T_{\text{ON}}$ beacons should be a time $t_4$ or at least a multiple of this time.
Specifically, $t_4$ is set according to {(\ref{eq5}), %and (\ref{eq8}) 
%we calculate $t_4$, 
where $t_{\text{sleep}}$ is predicted by the forecasting model based on} the data. 
%Then, %if the time predicted is below $1.5\times t_4$, the time between $T_{\text{ON}}$ of duration $t_1$ is set to $t_4$. On the other hand, if $t_{\text{sleep}}$ predicted by the forecasting model is above $1.5\times t_4$, then we calculate how many steps $n$ of duration $t_4$ fit in $t_{\text{sleep}}$, therefore, the time between $T_{\text{ON}}$ of duration $t_1$ is set to $n\times t_4$.
%\textcolor{blue}{If} \{$t_{\text{sleep}}/t_4$\} < 0.5, then $n$ = floor($t_{\text{sleep}}/t_4$),
%otherwise $n$ = ceil($t_{\text{sleep}}/t_4$), where \{\} denotes fractional part.
Notice {that even} when the proposed $t_{\text{sleep}}$ is not equal to $t_4$, the WuS parameters were prearranged between the coordinator and the MTDs. Therefore, in order to maintain the synchronism between the coordinator and each MTD, the time between $T_{\text{ON}}$ beacons should {be a multiple of $t_4$.}}
%\vspace{-4pt}
\section{Performance Metrics}\label{sec5}
In this section, we discuss two performance metrics, named energy consumption and mean delay, {which are} utilized to assess the performance of the proposed method and to compare it with {the benchmarks.} We find expressions for these metrics based on the Markov chain in Fig.~\ref{figure2.2} for a given WuS configuration. {Additionally, to confirm the accuracy of the predictor, we calculate the root mean square error (RMSE) between the estimated ($g_{i}$) and actual ($g_{o_{i}}$) inter-arrival values,} %The RMSE is obtained as follows
\begin{equation}
	\text{RMSE} = \sqrt{{\frac{1}{M}\sum_{i=1}^{M}\big(g_{i}-g_{o_{i}}\big)^{2}}},
\label{eq6}	
\end{equation}
{where $M$} is the sample size. % \rds{N já foi usado para outra coisa}. 
{Moreover}, the fitting performance is validated and tested using the R metric{, defined as }  
\begin{equation}
	{\text{R} = 
	\sqrt{1-\frac{1}{M}\sum_{i=1}^{M}{{\Big(1- \frac{g_{i}}{g_{o_{i}}}\Big)^{2}}}}.}
\label{Rsquared}	
\end{equation}  

\noindent %that represents the ratio between the variations experienced in the data that are not explained by the predictor and the total variations (sample size). 
This statistic measures how successful the fit is in explaining the variation of the data. Therefore, it represents the correlation between the actual values and the predicted ones. In general, the higher the {R metric }is, the better the model fits the data.

\subsection{Energy Consumption}
{In this paper, we adopt a simplified model based on~\cite{rostami2019wake} that refers only to the energy consumption of the RFI, i.e., $\text{PW}_{4}=0$. Based on Fig.~\ref{figure2.1}, which shows the power consumption levels ($\text{PW}_{i}$) for every WuS state $S_{i}, \ i = (1,2,3,4)$, the mean power consumption ($\overline{\text{PW}}$) is calculated as %\footnote{This equation was resize with algebraic transformations by multiplying by 2 both, the numerator and denominator, for aesthetic purpose in order to fit it in the column.}: 
\begin{equation}
	\overline{\text{PW}} \!=\! \frac{\sum\limits_{i=1}^{4}\! p_{i}t_{i}\text{PW}_{i}\! +\! \frac{1}{2}\big( p_{1}P_{12}t_{u}\text{PW}_{2}\!+\!p_{3}P_{34}t_{pd}\text{PW}_{3}\big)}{ p_{1}P_{12}t_{u}+p_{3}P_{34}t_{pd} + \sum\limits_{i=1}^{4} p_{i}t_{i}},
\label{eq7}	
\end{equation}
where $p_{i}$ denotes the probability of being in state $S_i$. Thus, by exploiting $p_{i} = \sum_{j=1}^{4} p_{j}P_{ji}$, $\sum_{i=1}^{4} p_{i} = 1$, and the semi-Markov chain in Fig.~\ref{figure2.2}, we obtain
%
%\begin{subequations}
\begin{align}
    %& p_{i} = \sum\limits_{j=1}^{4} p_{j}P_{ji},\ \text{and}\ \sum_{i=1}^{4} p_{i} = 1.\\
    p_{1} = p_{4} = \frac{(q-1)^2}{L},\ 
    p_{2} = \frac{P_A}{L},\  
    p_{3} = \frac{P_A(1-q)}{L}.
    %& L = 2P_A - 4q - qP_A + 2q^2 + 2.
    \label{Stable_state}
\end{align}
%\end{subequations}
where $L = 2P_A - 4q - qP_A + 2q^2 + 2$.}

%\textcolor{red}{
%\begin{align}
%p_{i} = \sum\limits_{j=1}^{4} p_{j}P_{ji},\ \text{and}\ \sum_{i=1}^{4} p_{i} %= 1. \label{Stable_state}%\\
%\end{align}
%
%The power consumption is normalized to mW/TTI. Then, 
The time the device spends in each state, multiplied by its corresponding power consumption, over the total time, outputs the mean power consumption. {Moreover,} the time to reach the state could be dismissed except for the start up time ($t_{\text{u}}$) and the power down time ($t_{\text{pd}}$), {whose values are %long
large enough to be considered.} The power consumption in these transitions is approximated, without losing generality, by their triangle geometrical format (see Fig.~\ref{figure2.1}), leading to the 1/2 %coefficient, i.e., for power consumption during 
coefficient in~(\ref{eq7}). %For instance, the power consumption during $t_{\text{u}}$ is $\text{PW} = %\frac{1}{2} 
%1/2 \times t_{\text{u}} \times (\text{PW}_2 - \text{PW}_4)$.
\subsection{Delay Constraints}

%For this sake, we need to obtain an expression to calculate the mean packet delay based on the WuS and traffic model. 
We assume that a packet experiences delay only when the situation that generates the detection of a triggering event finds the MTD 
%it is received while the system is 
in sleep $(S_{4})$ or WRx-On $(S_{1})$ state. This is because in active $(S_{2})$ and inactivity timer $(S_{3})$ states{, the} %the 
exchange of information with the coordinator is immediate. %So
Then, $\overline{D}$ can be estimated as 
\begin{equation}
	\overline{D} \!=\! p_{4}P_{A}\bigg( 3(t_\text{u}\!+\!t_{1})+%\int_{0}^{t_{4}}t\partial t 
	\frac{t_4^2}{2}\!+\! \sum_{n=1}^{\infty} p_\text{md}^n t_{4}\bigg)\!+\!t_\text{mac}.
\label{eq8}	
\end{equation}

\noindent Packets experience delay in 3 cases:
\begin{enumerate}
    \item a packet is intended to be delivered while the RFI of the {MTD} is at $S_1$. In this case, the packet experiments a delay $t_\text{u} + t_1${;\footnote{{In practice, $t_\text{u}$ dominates here due to the very %low duration
    small values of $t_1$.}}}
    \item a packet is buffered on hold while the {RFI of the MTD} is at $S_4$ state. Then, the {experienced} delay  is {given by} $t_\text{u} + t_1 + {\textstyle \int_{0}^{t_{4}}}t\partial t$ =  $t_\text{u} + t_1 + t_4^2/2$;
    \item the {MTD} does not receive the PO scheduled for information exchange, {thus, there is a} miss-detection. In this case, the device has to wait until the next PO, {and} the packet experiences a delay $t_u + t_1 + \textstyle \sum_{n=1}^{\infty} p_\text{md}^n t_{4}$. Herein, $p_\text{md}$ represents the miss detection probability and $n$ is the number of lost POs. 
\end{enumerate}
{The mean delay also includes a term $t_\text{mac}$ that comprises the delay introduced by the MAC scheme aiming at avoiding/solving collisions.} We assume that the communication requirements could be met with a suitable MAC algorithm.%any of the MAC algorithms proposed for 5G \rds{que são quais? não é melhor dizer "a suitable MAC algorithm"?}.
%, therefore, $t_{mac}$ represents the delay due to the MAC protocol.

\subsection{Power Saving Factor}

Finally, we calculate the relative power saving ($\eta$) regarding {a} benchmark, thus, it quantifies the amount of energy that can be saved by implementing our proposed scheme. This power saving factor is calculated {as} %follows
\begin{equation}
	\eta = \frac{\overline{\text{PW}}_{b} - \overline{\text{PW}}}{\overline{\text{PW}}_{b}},
\label{eq9}	
\end{equation}where $\overline{\text{PW}}_{b}$ represents the power consumption of the benchmark scheme and is calculated using~\eqref{eq7} with the values %of $p_i$ and $P_{ij}$ 
respective to the benchmark.% \rds{que é calculado como?}
%(in this case we use~\cite{rostami2019wake} as it is the one with the best performance of the two mechanisms proposed for comparison). 
%This metric provides a better understanding of the improvement in efficiency by implementing our model with respect to the benchmarks.

\section{Numerical Results}\label{sec6}

In this section, we evaluate the performance of the proposed method. 
Without losing generality, we assume a negative exponential function %$\left(f(d)=e^{-d}\right)$ %(see Table~\ref{table1})
to model the influence of events on {the traffic of the MTDs}, i.e., $p(d)=e^{-d}$.
The MTDs are deployed with density $\lambda_{M} = 10^{-1} (\text{MTDs}/m^{2})$, and TTI is assumed equal to 1 ms. 
All devices are at the state \textit{I} at the beginning of the simulation.
These parameters are set for all the  numerical results unless specified otherwise.
%All algorithm programming and data analysis was carried using MatLab~\cite{MATLAB}.  }

\subsection{Benchmarks}

%In this sense, we aim to implement the system model at 3 moments, where the coordinator has a different configuration in each one, the proposed FWuS method and 2 others that perform as benchmarks. 
As benchmarks, we resort to the WuS scheme in~\cite{rostami2019wake}, 
%with a standard WuS behavior (see Fig.~\ref{figure2.1}), 
and the optimized DRX-based reference mechanism in~\cite{ramazanali2016tuning}. 
The scheme in~\cite{rostami2019wake} models a standard {WuS behavior} ({see Fig.~\ref{figure2.1} b))}. 
In this scheme, 
the time interval between WRx-On states ($S_1$) is {predefined %in relation to the BS
 in} a static and invariable way. 
%Section~\ref{sec31} shows This, depending on the parameters delay and average packet arrival rate as shown in . 
%The intention of our proposal is to eliminate different WRx-On states %in which the probabilities of receiving WuS signals are almost zero%
%based on the traffic behavior throughout the session. Thus, enabling the RFI of the device to save energy by dynamically adapting to traffic conditions. 
The configuration values of the WuS scheme %are obtained in
come from~\cite{rostami2019wake}.
Meanwhile, the
%The 
basic operation and representative power consumption behavior of the DRX-enabled conventional cellular module~\cite{ramazanali2016tuning}  is shown in {Fig.~\ref{figure2.1} a)}. Observe that the RFI remains ON for a time $t_\text{on}$, during which the MTD is waiting  to receive information packets. If during this time no packets are received, the RFI goes to the OFF state, completing the DRX cycle. This cycle is repeated {until a data packet is received, after which} 
%as long as no packets arrive. On the contrary, if packets are received, 
the system goes to the active-decoding state with duration $t_{2}$. {Then, the inactivity timer is activated} for a time $t_{3}$. If packets are received in this state, the MTD returns to the active-decoding {state% \rds{isso gera confusão, pois A state neste trabalho é outra coisa, ou não?}
, otherwise, it goes to the OFF state. In every case, the process is repeated.} %, and the process is repeated. 

In state \textit{A}, the consumption is higher than in the other states ($\text{PW}_2$). In the inactivity timer state and in the ON state of the DRX cycle, the consumption is equal to $\text{PW}_3$. In the OFF state of the DRX cycle, the consumption is  $\text{PW}_4$. Note that the DRX cycle consists of a time in the ON state plus a time in the OFF state. The DRX configuration values are taken from~\cite{ramazanali2016tuning}.

%To perform the numerical evaluations, we use the power consumption profile in Fig.~\ref{figure2.1} b) for~\cite{rostami2019wake}, the profile in Fig.~\ref{figure2.1} c) for our proposed model, while Fig.~\ref{figure2.1} a) shows the consumption profile of the DRX-mechanism~\cite{ramazanali2016tuning}.   
\subsection{Simulation Framework}\label{result2}
\begin{figure}[t!]
    \centering
	%\subfigure[RMSE values during training process.]%{0.49\textwidth}
    %    {
    %    \centering
        \centerline{\includegraphics[width=0.97\columnwidth]{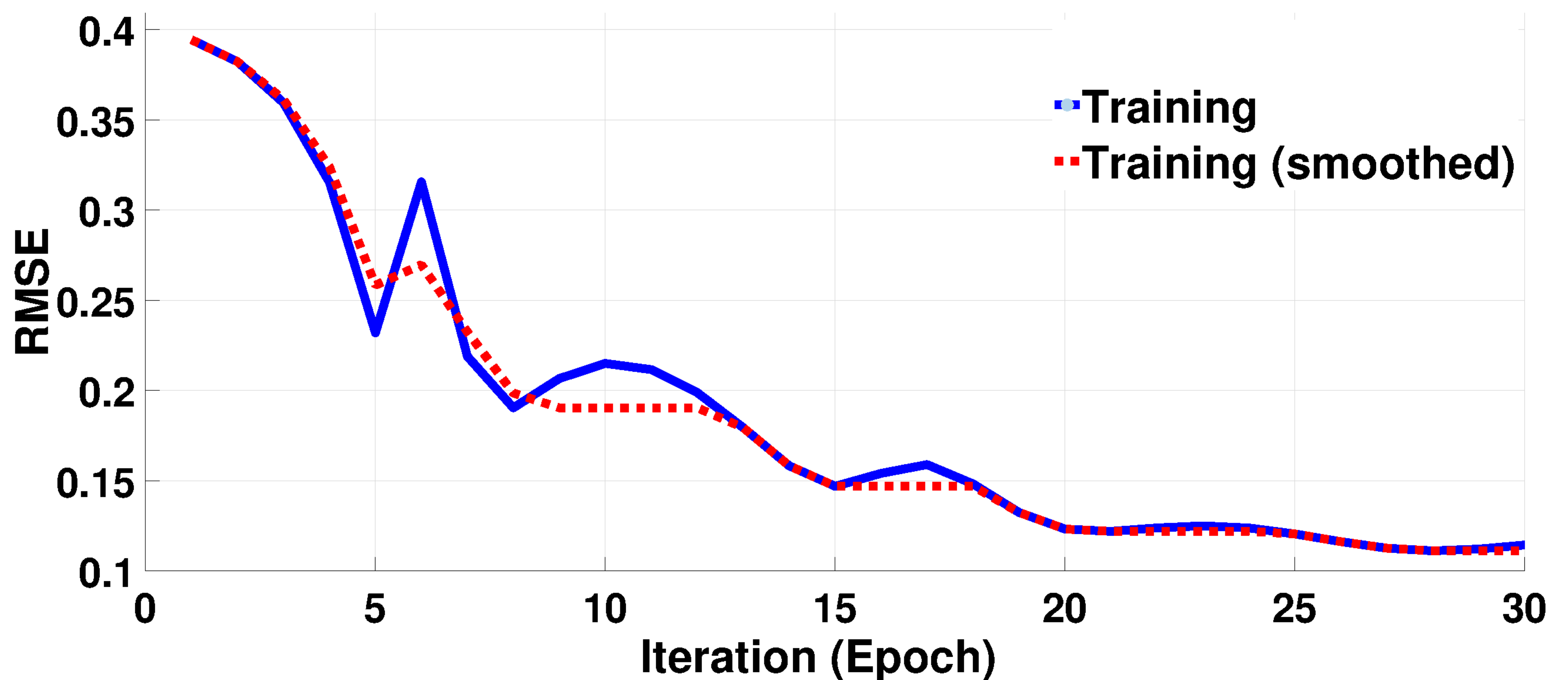}}
    %    }
        %\caption{RMSE values during training process.}
	%\\
	%\subfigure[Validation and Test processes.]%{0.49\textwidth}
    %    {
    %    \centering
        %\centerline{\includegraphics[width=0.65\columnwidth]{img/Validation_Test_1.pdf}}
        %\caption{Validation and Test processes.}
	%    }
	    %\centering{\includegraphics[height=2cm, width=0.3\columnwidth]{img/RBox.pdf}}
	   %\begin{minipage}{.55\columnwidth}
            \centerline{
            \includegraphics[%height=2.3cm, 
            width=1.02\columnwidth]{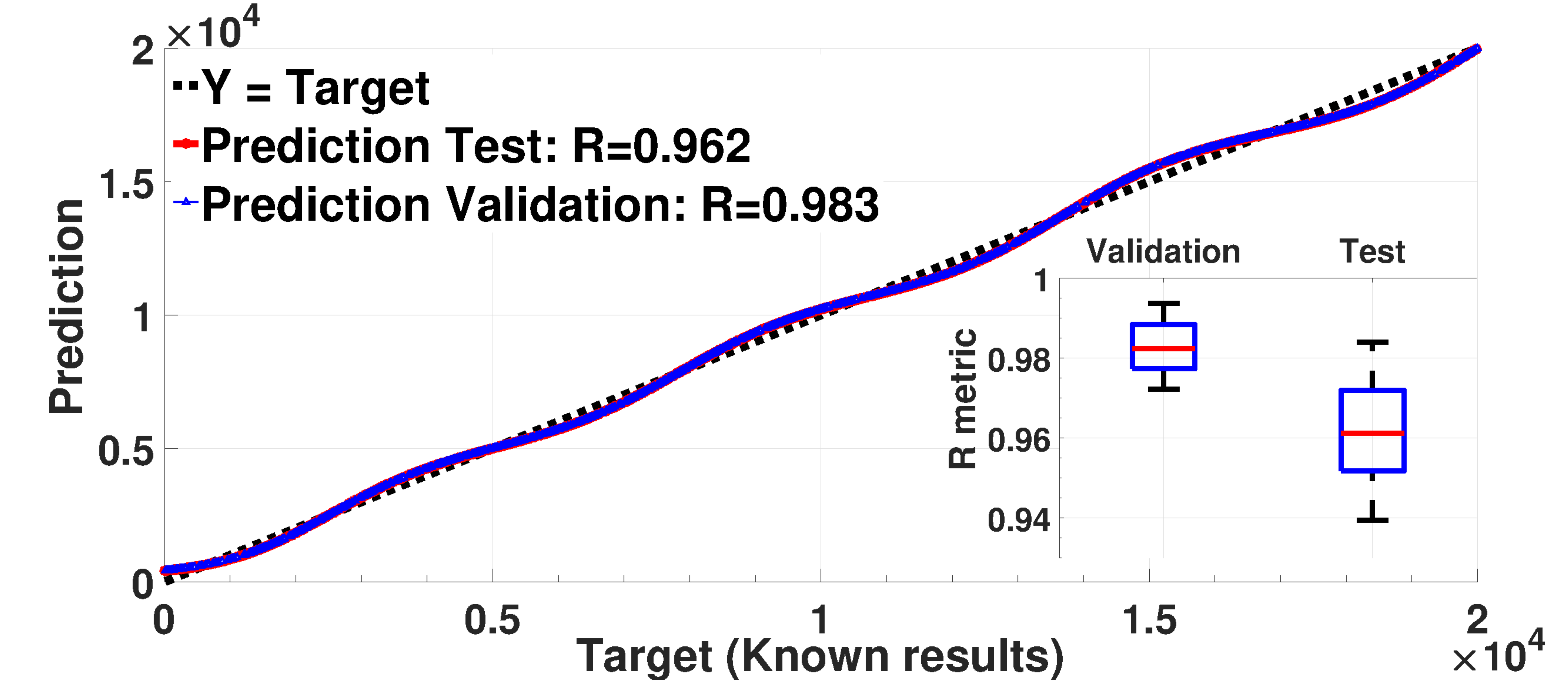}}
        %\end{minipage}%
        %\begin{minipage}{.45\columnwidth}
            %\centering
            %\includegraphics[height=2cm, width=\linewidth]{img/RBox.pdf}
        %\end{minipage}
    \caption{{Training and validation process of the forecasting model. a) RMSE values during training process for 30 epochs (top), b) validation and test processes (bottom). The latter shows the correspondence between predicted and known results{. The} dotted black line shows the ideal scenario where the predicted and known results match% (bottom)%, and c) statistics on the predictor accuracy, std = 0.0227 (bottom right)
    .}}
    \label{figure11}
\end{figure}
{We} use the traffic model in Section~\ref{sec32} to simulate several log traces of traffic received at the coordinator. In each log trace, the {deployment of the MTDs and 
%are deployed according to PPP $\Phi_{M}$, then the 
event epicenters is generated according to the corresponding PPP.} 
We consider the influence of each event epicenter {on the MTDs via $p(d)$. The latter influences $P_A$ in the traffic model through~\eqref{eq3}.}
%Without losing generality, in this paper we assume a negative exponential function $\left(f(d)=e^{-d}\right)$%(see Table~\ref{table1})
%to model the influence of events on device traffic.
The %supervised alarm \rds{acho que não usamos mais este termo} 
events are independent among one another.
%The MTDs are deployed as a network with a density $\lambda_{M} = 10^{-1} (nodes/m^{2})$.
These traffic data are then used as input of the LSTM network to train the forecasting model%, 
.
{Specifically,} 70$\%$ of the sample data is used to perform training, 15\% for the validation {set, while} the remaining 15\% for the testing set. %For training process a total of $9.8\times 10^4$ information packets were processed, while $2.1\times 10^4$ were used for validation and testing.} %while the remaining 30$\%$ is used for validation and testing. 
%As a result, we obtain a trained model capable of predict the inter-arrival time between information packets. 
%For training process a total of $9.8$×$10^4$ information packets were processed, while $4.2$×$10^4$ were used for validation and testing ($2.1$×$10^4$ for each one). In total, $1.4$×$10^5$ information packets are taken into account.}
For the training, validation, and testing process, $9.8$×$10^4$, $2.1$×$10^4$, and $2.1$×$10^4$ information packets were respectively processed, thus making a total of $1.4$×$10^5$. {The LSTM architecture is configured with} 1 hidden layer (100 Neurons), `Initial Learn Rate' equal to 0.0001, {RMSE} loss function, maximum 50 epochs~\cite{eramo2022application, chen2021expressway}, and using the Adam optimizer~\cite{sharma2017adam}. {Note that the} LSTM's learning algorithm is local in space and time, while its computational complexity per time step and weight is $O(1)$~\cite{gers2000learning}, thus, leading to a system complexity $O(2\times 100)$.

{The traffic forecasting model is then used to optimize the WuS parameters. 
Specifically, 
%In a second moment, 
the coordinator is configured with a standard WuS behavior (see {Fig.~\ref{figure2.1} b)}). Values of WuS parameters are obtained from~\cite{rostami2019wake}, while the value for $t_4$ is re-calculated using the framework in~\cite{rostami2019wake} for our system and traffic models aiming to be fair in the comparison. Lastly, we provide the coordinator RFI with the optimized DRX-based reference mechanism in~\cite{ramazanali2016tuning}, as suggested by ~\cite{rostami2019wake}. The DRX cycle values are adapted according to our traffic model. In this case, $t_\text{on}$ is established equal to 1 TTI in order to save energy, while $t_\text{u}$, $t_\text{pd}$, $t_2$ and $t_3$ have common values for each method.} 
%All algorithm programming and data analysis was carried using MatLab~\cite{MATLAB}. 

\begin{table}[t!]
\caption{Simulation Parameters}
\label{table1}
\centering
%\processtable{Parameters used in the simulation.\label{table1}}
\begin{tabular}{lll}
\hline
Parameter                                		& Value                                & Ref.                                         	\\
\hline
\\
$t_{1}$                & 1/14 (TTI)                         & \cite{rostami2019wake}      	\\
$t_{2}$     	& 1 (TTI)                              & \cite{rostami2019wake}      	\\
$t_{3}$               	& 1 (TTI)                              & \cite{rostami2019wake}      	\\
%$t_{4}$ (Sleep time)                     	& (\ref{eq8})     					&                                              		\\
$t_\text{u}$                  	& 15 (TTI)                            & \cite{rostami2019wake}     	 	\\
$t_\text{pd}$          	& 10 (TTI)                            & \cite{rostami2019wake}      	\\
$\overline{D}$     	& 30 ms                                     	&                                              		\\
$PW_{1}$ 		& 57 mW       & \cite{rostami2019wake}      	\\
$PW_{2}$ 		& 935 mW    	& \cite{rostami2019wake}      	\\
$PW_{3}$		& 850 mW    	& \cite{rostami2019wake}      	\\
$PW_{4}$ 		& $ \approx 0$ mW  	& \cite{rostami2019wake}      \\
$\lambda_{E}$      & $10^{-5}, 10^{-4}, 10^{-3}, 10^{-2}$ & \cite{thomsen2017traffic}   \\
$\lambda_{M}$      & $10^{-1}$ (MTD/$m^2$) & \cite{thomsen2017traffic}   \\
$p(d)$             	& $e^{-d}$                       	& \cite{thomsen2017traffic}   	\\
$q$			& [0.1, 0.9]   	&                                              		\\
$t_\text{on}$                       						& 1 (TTI)        & \cite{ramazanali2016tuning} 	\\
$p_\text{md}$    			& 0.01           	& \cite{rostami2019wake}      	\\
$p_{f}$      				& 0.1             	& \cite{rostami2019wake}      	\\ 
\hline
\end{tabular}
\label{table1}
\end{table}

{Table~\ref{table1} 
summarizes the parameters used in the simulation process. {We calculate $t_4$ (sleep time) using~\eqref{eq5} and~\eqref{eq8}.}
{All simulations were carried out in $\text{Matlab}^{\text{\textregistered}}$~\cite{MATLAB}.}
} {The
illustrated curves are the result of Monte Carlo simulations
over 150 runs, where the position of the MTDs and the
events’ epicenter are randomly distributed in each run.}

\subsection{{Prediction Accuracy}}\label{result22}
\begin{figure} [t!]
            \centering
                \includegraphics[width=0.97\columnwidth]{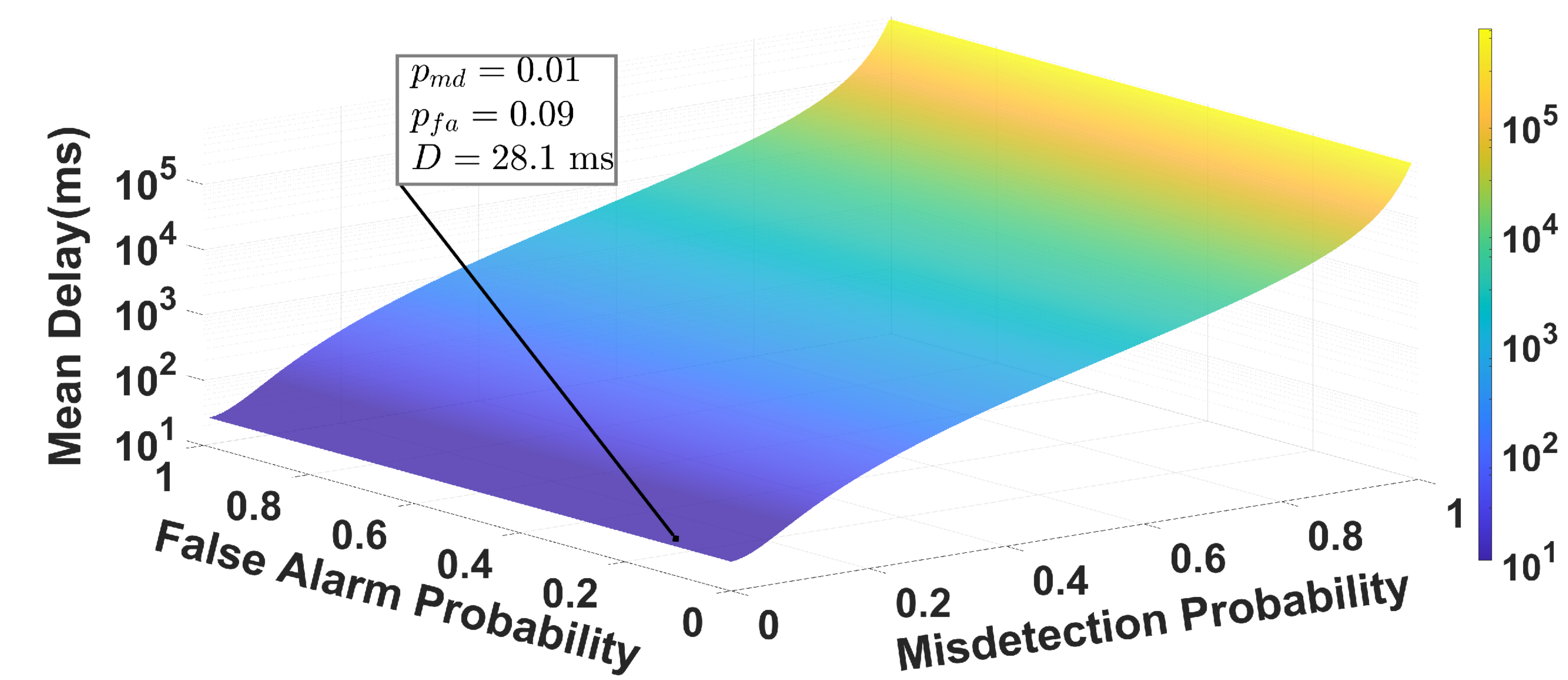}
                \includegraphics[width=\columnwidth]{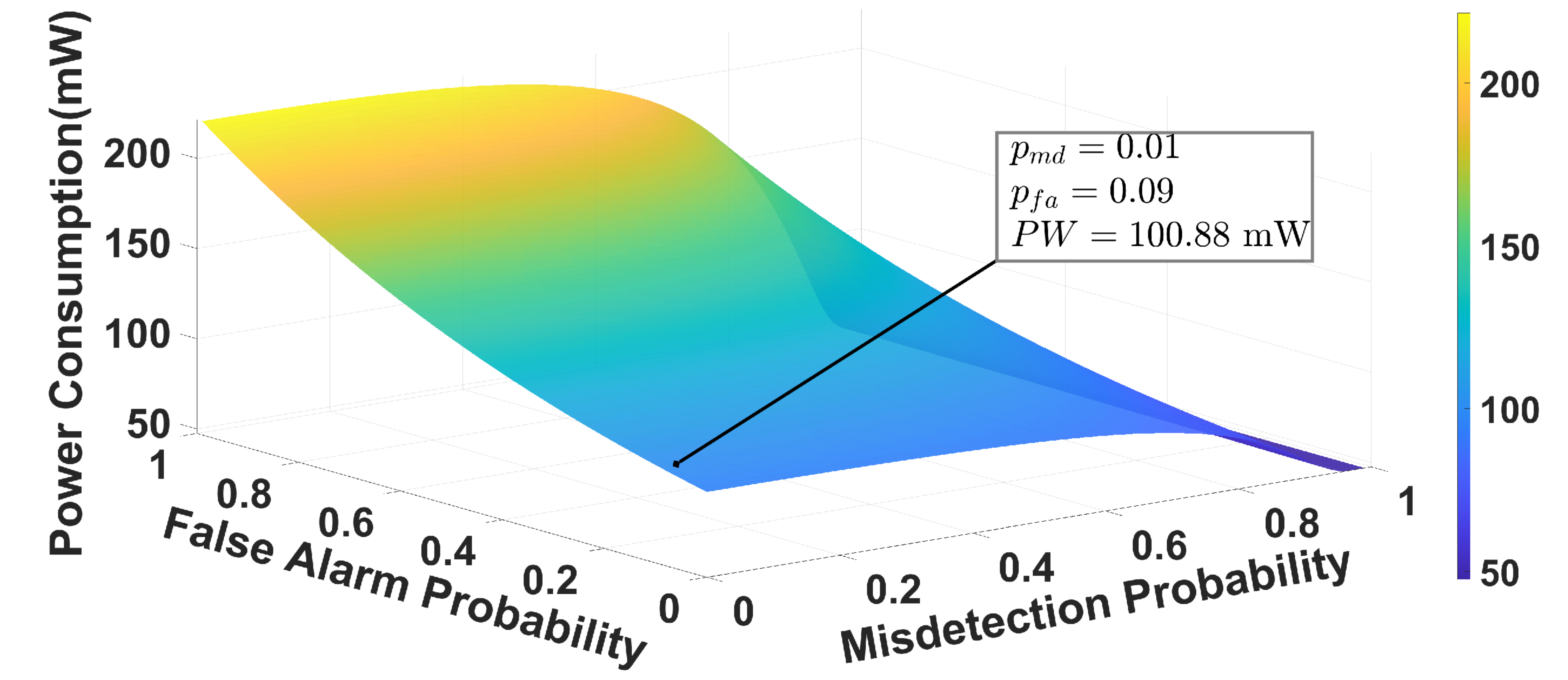}
                \caption{{System behavior regarding variations in miss-detection and false alarm probabilities{: a)} mean delay (top) and b) power consumption (bottom). {The marker represents the operation point} of our forecasting model.}}
                \label{Answer30}
        \end{figure}

Fig.~\ref{figure11} a) illustrates the performance of the forecasting training process{. Note} that after 30 epochs, {the RMSE value remains approximately constant at around 0.1.} In the validation and test processes{, the} obtained R values % \rds{a notação precisa ser revista, temos R state, R values, que são coisas diferentes}
were above 96$\%$, demonstrating high accuracy to adapt to incoming data. 
%\textchange{Fig.~\ref{figure11} shows the training, validation and testing processes. The sets were selected in this way, 70\% of the data for the training set, 15\% for the validation set and the remain 15\% for the testing set. For training process a total of $9.8\times 10^4$ information packets were processed, while $2.1\times 10^4$ were used for validation and testing.} 
{In addition, Fig.~\ref{figure11} b) shows the statistics on the accuracy of the predictor over 150 runs. The standard deviation for validation and test processes are 1.09\% and 2.27\%, respectively%Fig.~\ref{figure11} b) collects the mean value
.}
{%Confusion table for the forecasting model is also provided in Table~\ref{Answer19}. 
The obtained confusion table is as follows
\begin{align}
        %\begin{table}[h!]
        %\caption{{Confusion Table}}
        %\label{confusionp}
       % \centering
        \begin{tabular}{lll}
        \hline
            &\multicolumn{2}{c}{Prediction}\\
        WRx-On  & True                                & False                    
        \\
        \hline
        Positive	&91.2\%   &8.8\%\\
        Negative    &98.7\%  &1.3\%\\
        \hline
        \end{tabular}\nonumber
        %\label{Answer19}
     %   \end{table}}
\end{align}
Notice that miss-detection probability is actually the false negative probability{, while} the false alarm probability is actually the probability of taking a false positive. {Then, by adjusting these values{, we} are indirectly adjusting the accuracy of the predictor model. In this case, less than 10\% (8.8\%) of the energy saving opportunities are missed, while the risk for late activation is just 1.3\%.
Note that a deeper LSTM architecture or more training epochs would enhance the prediction accuracy but it would increase complexity and time of training as well.}
%LSTM's learning algorithm is local in space and time; its computational complexity per time step and weight is $O(1)$~\cite{gers2000learning}. Since we built a system with 1 hidden layer, the complexity is obtained as $O(m_1+m_1)$, where $m_1$ is the amount of neurons in the hidden layer.
%
%
\subsection{{Delay and Power Consumption Performance}}\label{result3}
Fig.~\ref{Answer30} a) and b) illustrate the required delay and power consumption{, respectively,} as a function of predictor's accuracy. The latter is given in terms of miss-detection probability (probability that the MTD misses a %page occasion or wake up signal
PO or WuS) and the false alarm probability (probability that the MTD receives a %page occasion or wake up signal 
PO or WuS needlessly, i.e., when there is no information to transmit/receive). As observed in Fig.~\ref{Answer30} a), the false alarm probability has no impact on the delay, while the miss-detection probability  increases less than 15\% (4.3 ms)  and 5\% (1.4 ms) for values below 0.1 and 0.05, respectively. {As for the power consumption, Fig.~\ref{Answer30} b) evinces the tremendous impact of the false alarm probability, while decreasing} the miss-detection probability below 0.4 has almost no impact}. 
 %Moreover, Fig.~\ref{Answer30} b) shows the variation in power consumption. Different from the false alarm probability, which impacts heavily the power consumption performance, decreasing the miss-detection probability below 0.4 has almost no impact%. on it. %However, notice that for small false alarm values (below 0.2) the variation in power consumption can be disregard. 
In this paper, the maximum values of miss-detection and false alarm probabilities, which serve as constraints to the forecasting model, are adjusted to 0.01 and 0.1, respectively.  %In fact, although values of misdetection below 0.1 do not have almost impact on the mean delay, values above 0.01 could affect the instantaneous delay. 
% that miss-detection probability is the false negative probability of the model and the false alarm probability is the probability of taking a false positive. Then, by adjusting these values, we are indirectly adjusting the accuracy of the predictor.
%Delay and power consumption values are normalized regarding the values of each of them for zero misdetection and false alarm probability, i.e., for a value `x' of misdetection probability and a value `y' of false alarm probability, the power consumption $PW_{x,y} = n\times PW_{0,0}$, where $n$ is the value obtained at the  z-axis in the figures.

\begin{figure}[t!]
    \centering
    
        %\centerline{\includegraphics[width=2.1\columnwidth]{img/PW_New2.pdf}}
    \begin{subfigure}
	    \centering
	    %\centerline
	    \includegraphics[width=0.495\textwidth]{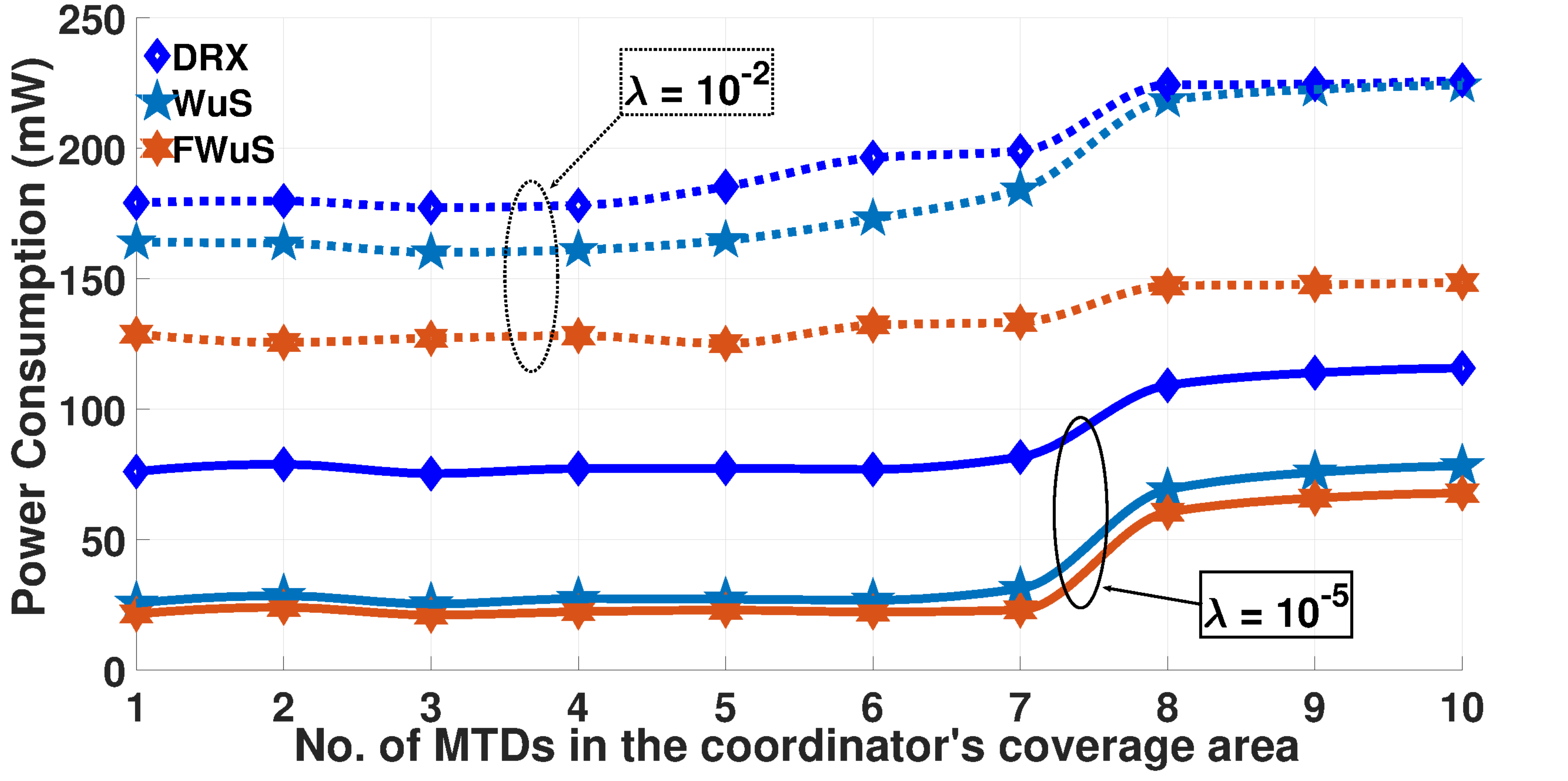}
	    %\caption{Energy consumption analysis as a function of the number of sensors in the coordinator's coverage area. We set $\lambda_{E}=10^{-5}$ and $q \in [0.1, 0.9]$. We obtain that $\Bar{\eta}\approx0.164$ holds independently of the number of MTDs.}
    %\label{figure7}
    \end{subfigure}  
    %\begin{subfigure}
	    %\centering
	    %\centerline
	    %\includegraphics[width=0.49\textwidth]{img/PW2.pdf}
	    %\caption{Energy consumption analysis as a function of the number of sensors in the coordinator's coverage area. We set $\lambda_{E}=10^{-4}$ and $q \in [0.7, 0.9]$. We obtain that $\Bar{\eta}\approx0.139$ holds independently of the number of MTDs.}
    %\label{figure8}
    %\end{subfigure}
    \begin{subfigure}
    	\centering
    	%\centerline
    	\includegraphics[width=0.495\textwidth]{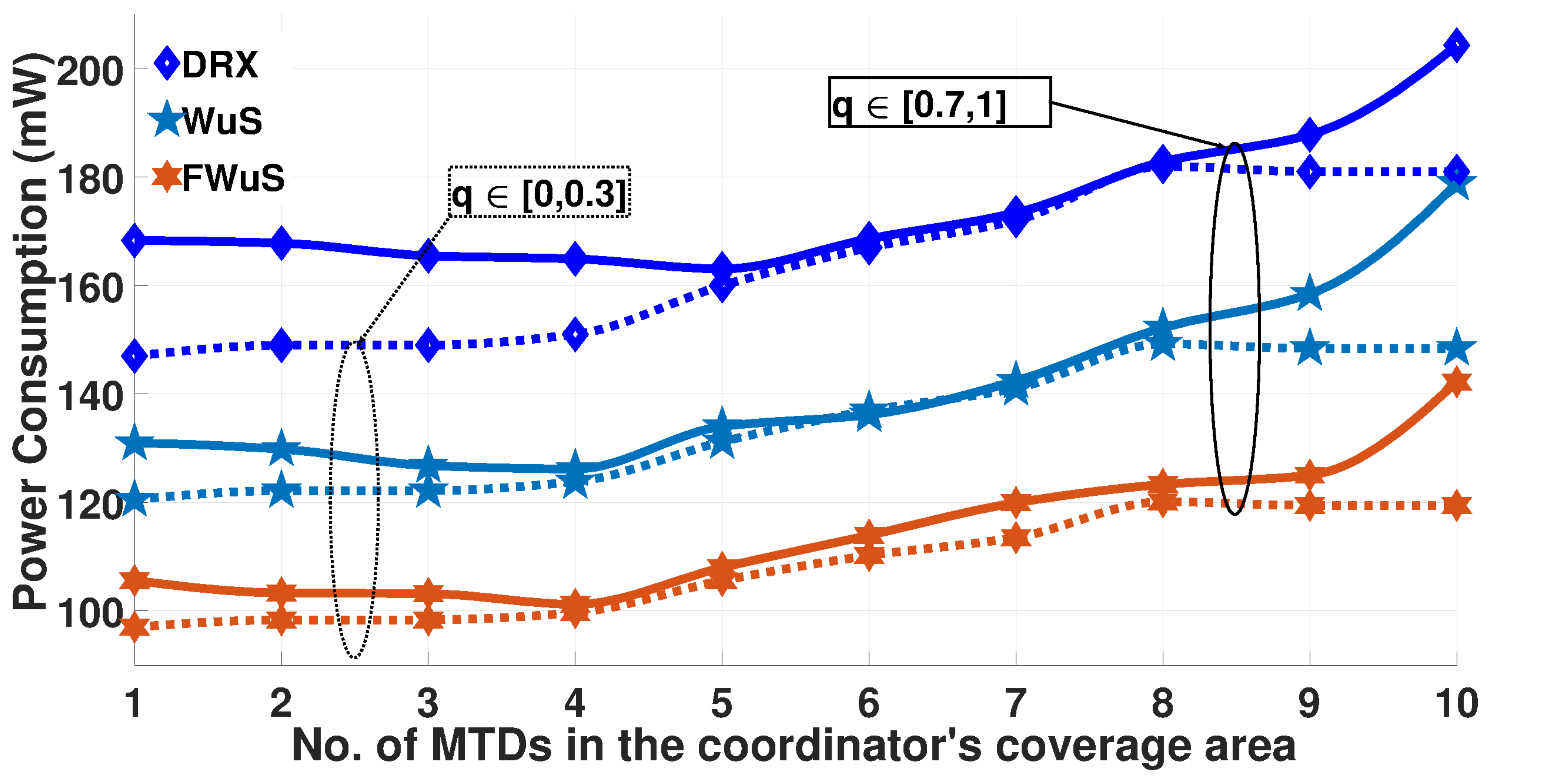}
    	%\caption{Energy consumption analysis as a function of the number of sensors in the coordinator's coverage area. We set $\lambda_{E}=10^{-3}$ and $q \in [0.4, 0.6]$. We obtain that $\Bar{\eta}\approx0.198$ holds independently of the number of MTDs.}
    %\label{figure9}
    \end{subfigure} 
    %\begin{subfigure}
    	%\centering
    	%\centerline
    	%\includegraphics[width=0.49\textwidth]{img/PW4.pdf}
    	%\caption{Energy consumption analysis as a function of the number of sensors in the coordinator's coverage area. We set $\lambda_{E}=10^{-2}$ and $q \in [0.1, 0.3]$. We obtain that $\Bar{\eta}\approx0.271$ holds independently of the number of MTDs.}
    %\label{figure10}
    %\end{subfigure}  
   
    %\caption{Energy consumption analysis as a function of the number of sensors in the coordinator's coverage area. a) $\lambda_{E}=10^{-5}$ and $q \in [0.1, 0.9]$, $\Bar{\eta}\approx16.4\%$ (top left), b) $\lambda_{E}=10^{-4}$ and $q \in [0.7, 0.9]$, $\Bar{\eta}\approx13.9\%$ (top right), c)  $\lambda_{E}=10^{-3}$ and $q \in [0.4, 0.6]$, $\Bar{\eta}\approx19.8\%$ (bottom left), and d) $\lambda_{E}=10^{-2}$ and $q \in [0.1, 0.3]$, $\Bar{\eta}\approx27.1\%$ (bottom right). The values of $\Bar{\eta}$ with respect to~\cite{rostami2019wake} holds independently of the number of MTDs.}
    \caption{Power consumption analysis as a function of the number of {MTDs} in the coordinator's coverage area. a) $\lambda_{E}\in\{10^{-5}, 10^{-2}\}$ and $q \in [0,1]$ (top), b) $\lambda_{E}={10^{-3}}$ and $q \in \{[0,0.3],[0.7,1]\}$,  respectively (bottom). %\textcolor{red}{..use 'MTDs' instead of 'sensors' in the x-axis label.}
    %Power consumption statistic analysis as a function of the number of sensors in the coordinator's coverage area. a) $\lambda_{E}=\{10^{-5}, 10^{-2}\}$ and $q \in [0,1]$ (top), b) $\lambda_{E}=\{10^{-3}\}$ and $q = \{[0,0.3],[0.7,1]\}$,  respectively (bottom).
    }
    \label{Power}
\end{figure}

{Fig.~\ref{Power} depicts the power consumption as a function of the number of MTDs for each scheme. %The results  are obtained via  Monte  Carlo  simulations  over  150  runs,  where  the position of the MTDs and the events’ epicenter are randomly distributed in each run.
In case of Fig.~\ref{Power} a), we consider event densities $\lambda_E\in\{10^{-5},10^{-2}\}$, while  each MTD is randomly assigned a value of $q \in [0,1]$. Note that our proposed scheme outperforms all the others, while the energy saving with respect to WuS \cite{rostami2019wake} becomes even greater as the traffic volume increases.} To generalize and better illustrate this, we adopt a different event density in Fig.~8 b), as well as different values for $q${.
Note} that the range of $q$ was varied with the aim of testing {more/less bursty traffic patterns}.
%Fig.~\ref{Power} b) shows the performance for $\lambda_E=10^{-4}$, with $q \in [0.7, 0.9]$, Fig.~\ref{Power} c) shows the performance for $\lambda_E=10^{-3}$, with $q \in [0.4, 0.6]$ and Fig.~\ref{Power} d) shows the performance for $\lambda_E=10^{-3}$, with $q \in [0.1, 0.3]$.\footnote[3]{Note that the $q$ values in the analysis of the aforementioned figures are also uniformly distributed among the MTDs.} 
In all the cases,  the energy consumption under the proposed scheme is considerably lower with respect to WuS \cite{rostami2019wake} and DRX \cite{ramazanali2016tuning}{.
%as the event density, and consequently event probability, increases. 
Interestingly, FWuS saves energy independently of the number of served MTDs under low bursty traffic, while DRX and WuS lead to the same power} consumption regardless of the type of traffic when serving 5 to 7 {MTDs.
Thus,} %the increasing pattern in power consumption varies as we vary %the value of 
the power consumption increases with both the number of served MTDs and 
$q$. For small $q$, the increase of the power consumption with the number of MTDs is smooth, while as $q$ increases, there might be sudden jumps in power {consumption.} %,
%the power consumption increases as increase the number of nodes. Meanwhile, as we increment the $q$ value (bursty pattern), the power consumption increases slightly with the number of nodes at first, and the jumps in power consumption occurs 
%each time for a greater number of nodes.
 
 {Note that since the power consumption becomes greater as the number of MTDs increases, the performance figures corresponding to 1 and 10 MTDs are relevant/representative. Hence, we characterize in more detail the power consumption performance for such numbers of MTDs in} {{Table~\ref{box}.} % and~\ref{box2}}}.
 Here, $\overline{\text{PW}}, \text{PW}_\text{max}$, and $\text{PW}_\text{min}$ represent the mean, maximum and minimum power consumption over the 150 Monte Carlo runs, while std is the standard deviation.} 

\begin{table}[t!]
\caption{{Power Consumption (\MakeLowercase{m}W) for 1 MTD (white) and 10 MTDs (lightgray)}}
\centering
%\processtable{Parameters used in the simulation.\label{table1}}
\label{box}
{\begin{tabular}{cccccc}
%\textchange{
\hline
\multicolumn{2}{c}{%\textchange{
Parameters%}
}      &\multicolumn{4}{c}{{Results}}\\
$\lambda_{E}$   & $q$		& {$\overline{\text{PW}}$} 	& $\text{PW}_\text{max}$ 	& $\text{PW}_\text{min}$ 	&std\\ \hline
\\
{$10^{-5}$}  &{$[0,1]$}	
&{21.423}		&{23.594}   	&{18.356} 		&{2.671}\\
\rowcolor{lightgray}
%{$10^{-5}$}   &{$[0,1]$}	
& &{67.738}     &{76.061}		&{59.164} 		&{8.602}\\
{$10^{-2}$}  &{$[0,1]$}	
&{127.723} 	&{142.143}	 	&{112.712} 		&{13.915}\\
\rowcolor{lightgray}%
& &{149.859} 	&{154.113}	 	&{134.215} 		&{14.212}\\
{$10^{-3}$}   &{$[0,0.3]$}	
&{97.023}	    &{110.468}		&{83.182} 		&{12.366}\\
\rowcolor{lightgray}%{$10^{-3}$}   &{$[0,0.3]$	}
& &{119.461}	    &{130.972}		&{107.041} 		&{11.667}\\
{$10^{-3}$}   &{$[0.7,1]$}	
&{105.603}	&{120.091}        &{93.889}   	&{9.659}\\
\rowcolor{lightgray}%{$10^{-3}$}   &{$[0.7,1]$	}
& &{142.239}	&{152.112 }       &{131.943 }  	&{9.011}\\
\hline
\end{tabular}}
\label{box}
%\footnotetext[4]{{Here, std is the standard deviation.}}
\end{table}

%\begin{table}[t!]
%\caption{{Power Consumption (\MakeLowercase{m}W) for 10 MTDs}}
%\centering
%\processtable{Parameters used in the simulation.\label{table1}}
%\label{box2}
%\begin{tabular}{cccccc}

%\hline
%\multicolumn{2}{c}{Parameters}      &\multicolumn{4}{c}{Results}\\
%$\lambda_{E}$   & $q$		& \textcolor{blue}{$\overline{\text{PW}}$} 	& $\text{PW}_\text{max}$ 	& $\text{PW}_\text{min}$ 	&std\\ \hline
%\\
%{$10^{-5}$}   &{$[0,1]$}	
%&{67.738}     &{76.061}		&{59.164} 		&{8.602}\\
%{$10^{-2}$}   &{$[0,1]$}	
%&{149.859} 	&{154.113}	 	&{134.215} 		&{14.212}\\
%{$10^{-3}$}   &{$[0,0.3]$	}
%&{119.461}	    &{130.972}		&{107.041} 		&{11.667}\\
%{$10^{-3}$}   &{$[0.7,1]$	}
%&{142.239}	&{152.112 }       &{131.943 }  	&{9.011}\\
%\hline
%\end{tabular}
%\label{box2}
%\footnotetext[4]{{Here, std is the standard deviation.}}
%\end{table}

\begin{figure}[t]
            \centering
            \includegraphics[width=0.82\columnwidth]{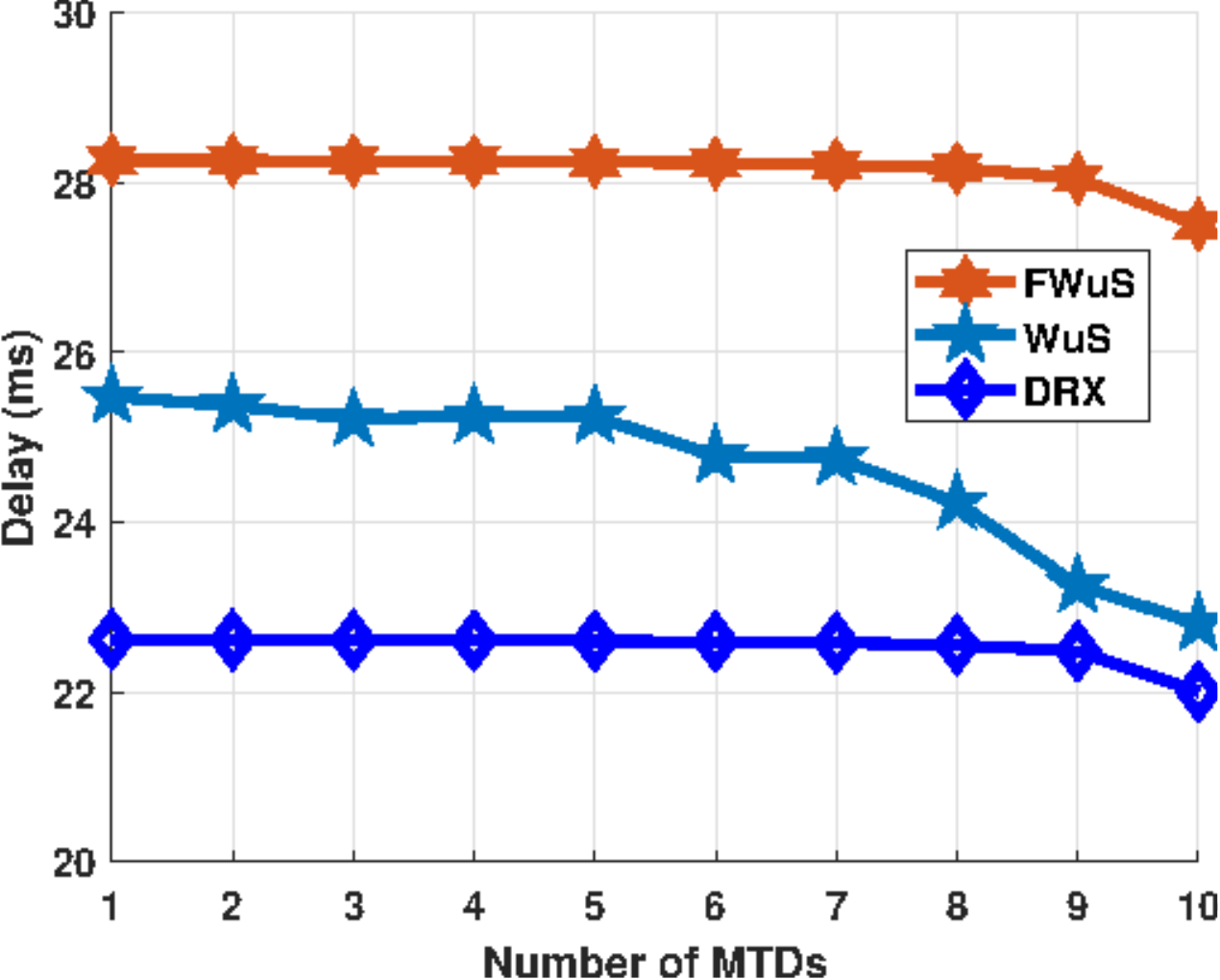}
            \caption{Delay analysis as a function of the number of {MTDs} handled by the coordinator. Maximum variations of $+1.4$ ms and $-2.1$ ms, with `std' equal to 1.1 ms when the number of {MTDs is} 10. %\textcolor{red}{..use 'MTDs' instead of 'sensors' in the x-axis label.}
            }
            \label{Delay}
        \end{figure}

{Fig.~\ref{Delay} illustrates} the mean delay experienced by each scheme. Notice that in all cases the delay is below the constraint of 30 ms, and the performance gap between our proposal and {WuS} \cite{rostami2019wake} is between 3 and 5 ms, while the gap regarding {DRX} \cite{ramazanali2016tuning} increases up to 6 ms. Notice that this figure illustrates the maximum mean delay when %varying the event density in the range %  $[10^{-5}$ $10^{-2}]$ 
{varying the event density between %these values 
\{$10^{-5}, 10^{-4}, 10^{-3}, 10^{-2}$\}}, thus representing the worst-case scenario for the mean delay.   

\subsection{Power Saving and Dynamic Adjustment}
%
%\textcolor{blue}{Herein, we} analyze the \textcolor{blue}{performance of FWuS under} changes in events \textcolor{blue}{density, thus,} changes in traffic \textcolor{blue}{behavior as} well as the training and adjustment capacity of the forecasting predictor.

{Table}~\ref{tab:2}
%Fig.~\ref{figure11} 
illustrates the relative power saving ($\eta$) regarding {WuS} \cite{rostami2019wake}. %showing 
In this case, we analyze the amount of energy that can be saved by implementing our proposed scheme instead of WuS, and quantify its dispersion via the {std metric.} Notice that we use {WuS} \cite{rostami2019wake} as it is more energy efficient than DRX~\cite{ramazanali2016tuning}. Evidently, the proposed FWuS saves much more energy than {WuS \cite{rostami2019wake}}, especially under high density of events and/or MTDs. Savings in energy consumption go from $10\%$ to around $%15 - 
20 \%$ for low-density event scenarios. Meanwhile, the savings increase up to almost $35\%$ for high-density event scenarios, thus guaranteeing  good performance despite harsh scenario dynamics.
\begin{table}[t!]
\caption{{Relative Power Saving ($\eta$)}}
\centering
%\processtable{Parameters used in the simulation.\label{table1}}
\label{tab:2}
\begin{tabular}{cccccc}

\hline
\multicolumn{2}{c}{Parameters}      &\multicolumn{4}{c}{Results}\\
$\lambda_{E}$   & $q$		& $\Bar{\eta}$ 	& $\eta_\text{max}$ 	& $\eta_\text{min}$ 	&std\\ \hline
\\
{$10^{-5}$}   &{$[0,1]$}	
&{16.4\%	}	&{26.3\% }  	&{12.6\% }		&{4.2\%}\\
{$10^{-2}$}   &{$[0,1]$}	
&{28.4.1\%} 		&{34.2\%}	 	&{22.9\% }		&{4.9\%}\\
{$10^{-3}$}   &{$[0,0.3]$}	
&{19.33\%}	&{21.3\%	}	&{16.8\% }		&{2.3\%}\\
{$10^{-3}$ }  &{$[0.7,1]$}	
&{19.5\%}		&{22.3\%	}	&{15.8\% }		&{2.7\%}\\
\hline
\end{tabular}
\label{tab:2}
%\footnotetext[4]{{Here, std is the standard deviation.}}
\end{table}
%\footnotetext[4]{{Here, std is the standard deviation.}}

\begin{figure}[t!]
	\centering
	\centerline{\includegraphics[width=0.98\columnwidth]{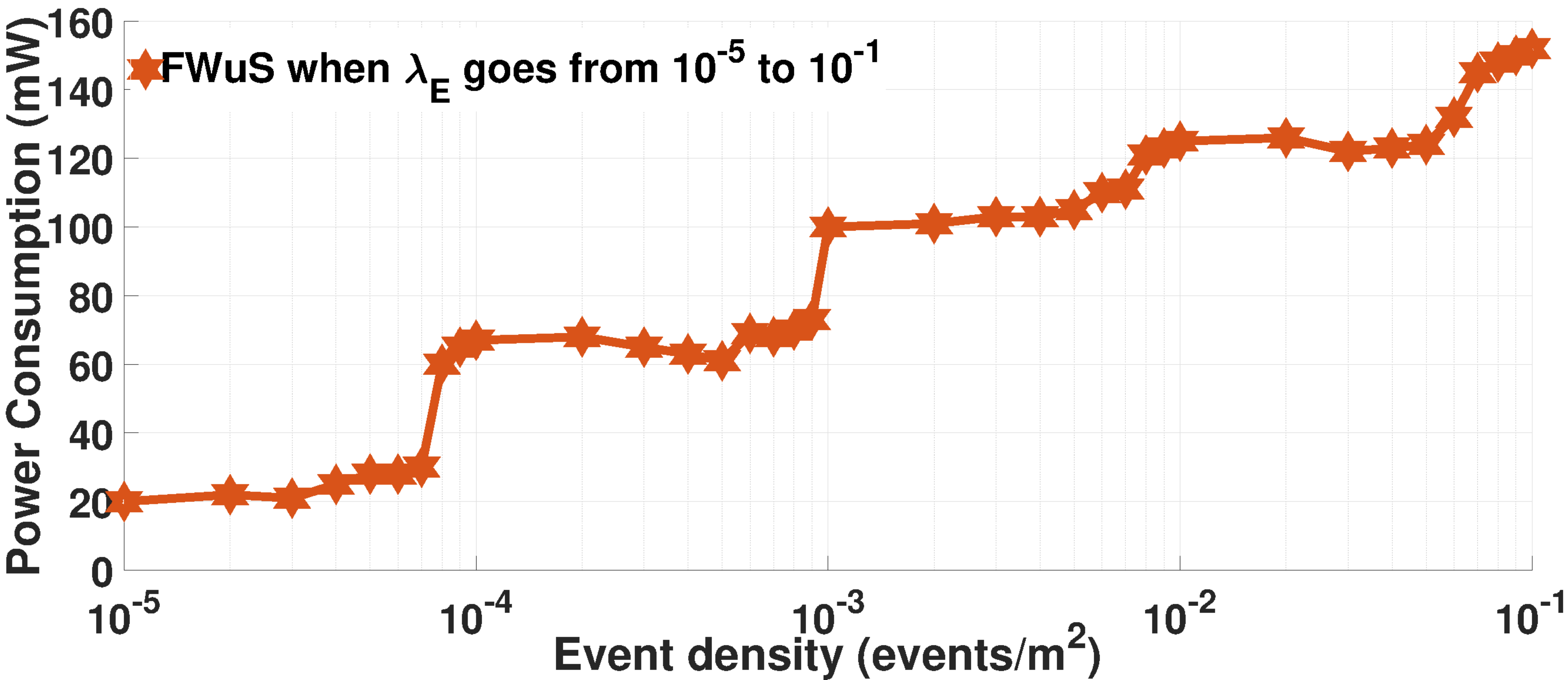}}
	\centerline{\includegraphics[width=0.98\columnwidth]{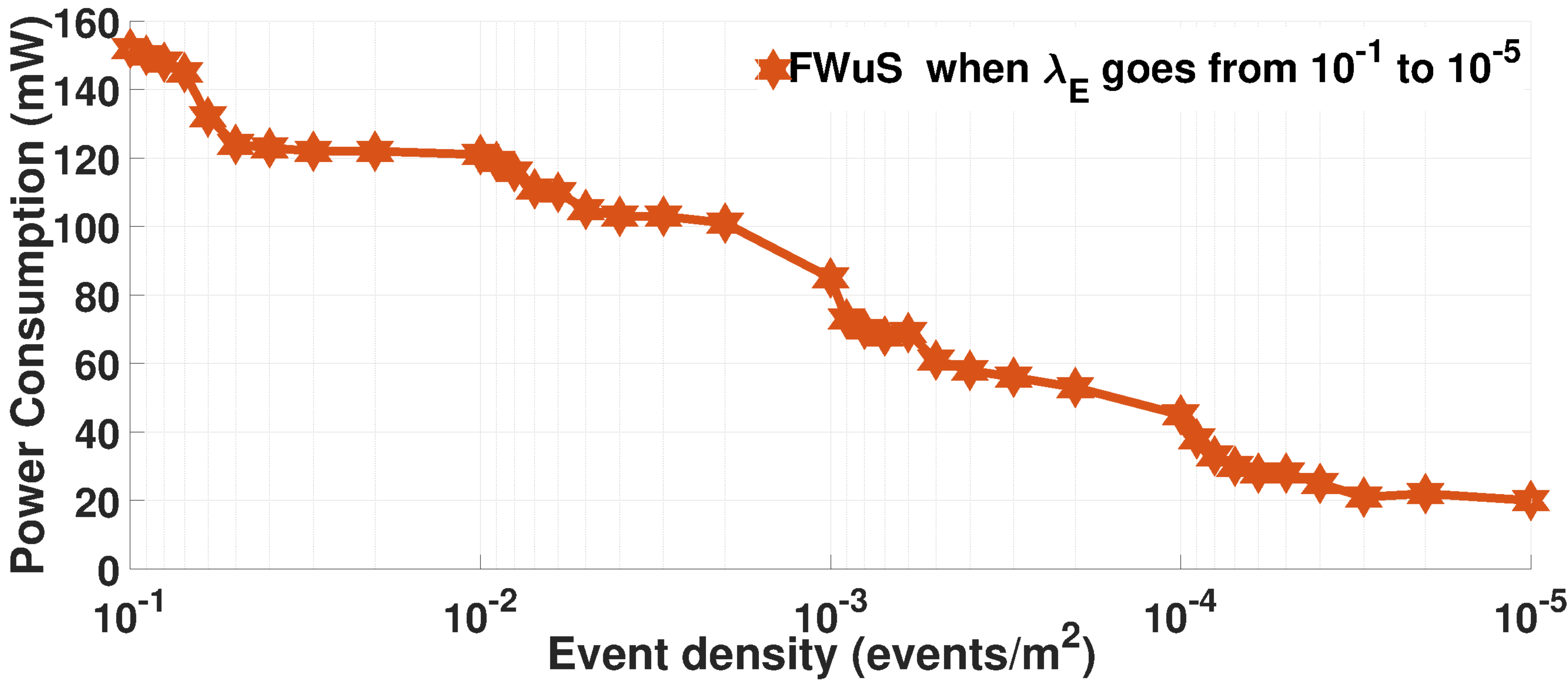}}
	\caption{{Energy consumption analysis when the event density changes dynamically throughout a period of time. a) $\lambda_{E} \text{ from } 10^{-5} \text{ to } 10^{-1}$ (top), and b) $\lambda_{E} \text{ from } 10^{-1} \text{ to } 10^{-5}$ (bottom).}}
\label{Density}
\end{figure}

{Fig.~\ref{Density} illustrates the performance of FWuS when the events density varies dynamically over a period of time. This is completely different from the approaches in sections \ref{result22} and \ref{result3}, where these parameters were completely deterministic. Specifically, events density increases from $10^{-5}$ to $10^{-1}$ in Fig.~\ref{Density} a), and decreases} from $10^{-1}$ to $10^{-5}$ in Fig.~\ref{Density} b). {Note that the} variations in event density change the activation probability of the MTDs, thus leading to changes in traffic patterns and challenging the online predictor. % \rds{não compreendi a partir de which while}.
{Nevertheless,} the proposed scheme is able to adapt itself to these  variations.  {Interestingly, the system adapts more} smoothly when the event density decreases since the scheme has more freedom to predict covered by the false alarm and miss-detection probabilities that adapt better to the decrease in traffic than to its increase. {Meanwhile,}  abrupt power consumption peaks may appear when the event density increases due to {the loss of multiple energy-saving} opportunities caused by the restriction imposed by the  miss-detection   probability target.

\section{Conclusions}\label{sec7} 
{In this paper, we considered an MTC network controlled by a single coordinator. We modeled the location of MTDs and event epicenters as independent PPPs, while considering  event-driven traffic patterns with geometrically distributed burst duration. %In the considered scenario, the information exchange is completely controlled by a 
We proposed an intelligent wake-up scheme (FWuS) to be used by the coordinator} to reduce {the} energy consumption in the MTDs and prolong their battery lifetime.
%Specifically, FWuS exploits an accurate traffic forecasting model that allows optimizing  the wake-up parameters, thus, favoring energy savings. 
{Specifically, FWuS exploits an accurate traffic forecasting model that optimizes the wake-up parameters, thus, favouring energy savings.}
The traffic forecasting relies on a simple {LSTM} neural network, which achieves prediction accuracy results above 96\% by monitoring the {MTC traffic patterns.}%Numerical results evinced that the proposed FWuS enables energy consumption reduction up to almost 35\% in relation to the competing scheme with the best performance, while the mean delay increases in only 2--6 ms \textcolor{blue}{but always} below the 30 ms constraint. 
{Numerical results evinced that the proposed FWuS enables energy consumption reduction up to almost 35\% to the competing scheme with the best performance, while the mean delay increases only 2--6 ms but is always below the 30 ms constraint.}
Moreover, the proposed mechanism was shown to be robust/adaptable to scenario/traffic with dynamically changing parameters, performing well under traffic variability.

{Some interesting research directions to pursue in the sequence are:
%
%\begin{itemize}
    %\item 
    i) considering/exploiting the traffic correlation between nearby MTDs that potentially detect the same events to avoid transmission of unnecessary redundant traffic, thus saving energy;
    ii) controlling the duration and restart procedure of the inactivity timer, e.g., via downlink control information, which seems appealing to reduce the active time; and
    iii) analyzing/comparing predictors' efficiency in terms of complexity and accuracy trade-offs, including more demanding scenarios such as those with `on device intelligence'.}
%\end{itemize}
%
Finally, acquiring and exploiting data from the industry, while validating the proposed mechanisms, is a goal for future work.

\bibliographystyle{IEEEtran}
\bibliography{bib.bib}

% Generated by IEEEtran.bst, version: 1.14 (2015/08/26)
\begin{thebibliography}{10}
\providecommand{\url}[1]{#1}
\csname url@samestyle\endcsname
\providecommand{\newblock}{\relax}
\providecommand{\bibinfo}[2]{#2}
\providecommand{\BIBentrySTDinterwordspacing}{\spaceskip=0pt\relax}
\providecommand{\BIBentryALTinterwordstretchfactor}{4}
\providecommand{\BIBentryALTinterwordspacing}{\spaceskip=\fontdimen2\font plus
\BIBentryALTinterwordstretchfactor\fontdimen3\font minus
  \fontdimen4\font\relax}
\providecommand{\BIBforeignlanguage}[2]{{%
\expandafter\ifx\csname l@#1\endcsname\relax
\typeout{** WARNING: IEEEtran.bst: No hyphenation pattern has been}%
\typeout{** loaded for the language `#1'. Using the pattern for}%
\typeout{** the default language instead.}%
\else
\language=\csname l@#1\endcsname
\fi
#2}}
\providecommand{\BIBdecl}{\relax}
\BIBdecl

\bibitem{mahmood2021machine}
N.~H. Mahmood, S.~B{\"o}cker, I.~Moerman, O.~A. L{\'o}pez, A.~Munari,
  K.~Mikhaylov, F.~Clazzer, H.~Bartz, O.-S. Park, E.~Mercier \emph{et~al.},
  ``{Machine type communications: key drivers and enablers towards the 6G
  era},'' \emph{EURASIP Journal on Wireless Communications and Networking},
  vol. 2021, no.~1, pp. 1--25, 2021.

\bibitem{9057670}
J.~Ding, M.~Nemati, C.~Ranaweera, and J.~Choi, ``{IoT Connectivity Technologies
  and Applications: A Survey},'' \emph{IEEE Access}, vol.~8, pp.
  67\,646--67\,673, 2020.

\bibitem{xia2019emerging}
N.~Xia, H.-H. Chen, and C.-S. Yang, ``{Emerging Technologies for Machine-Type
  Communication Networks},'' \emph{IEEE Network}, vol.~34, no.~1, pp. 214--222,
  2019.

\bibitem{akpakwu2017survey}
G.~A. Akpakwu, B.~J. Silva, G.~P. Hancke, and A.~M. Abu-Mahfouz, ``{A survey on
  5G networks for the Internet of Things: Communication technologies and
  challenges},'' \emph{IEEE Access}, vol.~6, pp. 3619--3647, 2017.

\bibitem{manogaran2020response}
G.~Manogaran, G.~Srivastava, B.~A. Muthu, S.~Baskar, P.~M. Shakeel, C.-H. Hsu,
  A.~K. Bashir, and P.~M. Kumar, ``{A response-aware traffic offloading scheme
  using regression machine learning for user-centric large-scale Internet of
  Things},'' \emph{IEEE Internet of Things Journal}, vol.~8, no.~5, pp.
  3360--3368, 2020.

\bibitem{YAN2022102601}
\BIBentryALTinterwordspacing
X.~Yan and M.~Ma, ``{A Privacy-Preserving Handover Authentication Protocol for
  a Group of MTC Devices in 5G Networks},'' \emph{Computers \& Security}, p.
  102601, 2022. [Online]. Available:
  \url{https://www.sciencedirect.com/science/article/pii/S0167404821004247}
\BIBentrySTDinterwordspacing

\bibitem{delay}
F.~Luo, X.~Sun, W.~Zhan, J.~Zou, and H.~Tan, ``{Distributed Delay Optimization
  of Machine-Type Communications in 5G Networks},'' in \emph{{Computing,
  Communications and IoT Applications (ComComAp)}}, 2021, pp. 138--143.

\bibitem{zanella2014internet}
A.~Zanella, N.~Bui, A.~Castellani, L.~Vangelista, and M.~Zorzi, ``{Internet of
  Things for smart cities},'' \emph{IEEE Internet of Things journal}, vol.~1,
  no.~1, pp. 22--32, 2014.

\bibitem{bayindir2016path}
R.~Bayindir, E.~Hossain, and S.~Vadi, ``{The path of the smart grid-the new and
  improved power grid},'' in \emph{International Smart Grid Workshop and
  Certificate Program (ISGWCP)}.\hskip 1em plus 0.5em minus 0.4em\relax IEEE,
  2016, pp. 1--8.

\bibitem{ratasuk2015recent}
R.~Ratasuk, A.~Prasad, Z.~Li, A.~Ghosh, and M.~A. Uusitalo, ``{Recent
  advancements in M2M communications in 4G networks and evolution towards
  5G},'' in \emph{18th International Conference on Intelligence in Next
  Generation Networks}.\hskip 1em plus 0.5em minus 0.4em\relax IEEE, 2015, pp.
  52--57.

\bibitem{Adrx}
J.~Wu, B.~Yang, L.~Wang, and J.~Park, ``{Adaptive DRX Method for MTC Device
  Energy Saving by Using a Machine Learning Algorithm in an MEC Framework},''
  \emph{IEEE Access}, vol.~9, pp. 10\,548--10\,560, 2021.

\bibitem{zhou2019online}
J.~Zhou, G.~Feng, T.-S.~P. Yum, M.~Yan, and S.~Qin, ``{Online Learning-Based
  Discontinuous Reception (DRX) for Machine-Type Communications},'' \emph{IEEE
  Internet of Things Journal}, vol.~6, no.~3, pp. 5550--5561, 2019.

\bibitem{li2020power}
Y.-N.~R. Li, M.~Chen, J.~Xu, L.~Tian, and K.~Huang, ``{Power Saving Techniques
  for 5G and Beyond},'' \emph{IEEE Access}, vol.~8, pp. 108\,675--108\,690,
  2020.

\bibitem{benbuk2021tunable}
A.~A. Benbuk, N.~Kouzayha, J.~Costantine, and Z.~Dawy, ``{Tunable,
  Asynchronous, and Nanopower Baseband Receiver for Charging and Wake-up of IoT
  Devices},'' \emph{IEEE Internet of Things Journal}, 2021.

\bibitem{shehab2020traffic}
M.~Shehab, A.~K. Hagelskj{\ae}r, A.~E. Kal{\o}r, P.~Popovski, and H.~Alves,
  ``{Traffic Prediction Based Fast Uplink Grant for Massive IoT},'' in
  \emph{IEEE 31st Annual International Symposium on Personal, Indoor and Mobile
  Radio Communications}.\hskip 1em plus 0.5em minus 0.4em\relax IEEE, 2020, pp.
  1--6.

\bibitem{9375479}
W.~Zhan, C.~Xu, X.~Sun, and J.~Zou, ``{Toward Optimal Connection Management for
  Massive Machine-Type Communications in 5G System},'' \emph{IEEE Internet of
  Things Journal}, vol.~8, no.~17, pp. 13\,237--13\,250, 2021.

\bibitem{dian2020lte}
J.~Dian and R.~Vahidnia, ``{LTE IoT Technology Enhancements and Case
  Studies},'' \emph{IEEE Consumer Electronics Magazine}, 2020.

\bibitem{mazloum2020interference}
N.~Mazloum and O.~Edfors, ``{Interference-free OFDM embedding of wake-up
  signals for low-power wake-up receivers},'' \emph{IEEE Transactions on Green
  Communications and Networking}, vol.~4, no.~3, pp. 669--677, 2020.

\bibitem{3gpp2019ts38}
3GPP, ``{TS38. 300: NR; NR \& NG-RAN overall description; Stage 2},'' 2019.

\bibitem{WuS}
\emph{{Draft Standard for Information Technology--Telecommunications and
  Information Exchange Between Systems--Local and Metropolitan Area
  Networks--Specific Requirements Part 11: Wireless LAN Medium Access Control
  (MAC) and Physical Layer (PHY) Specifications--Amendment 9: Wake-Up Radio
  Operation}}, Standard IEEE P802.11ba, Jun. 2020.

\bibitem{stepanova2020joint}
E.~Stepanova, D.~Bankov, E.~Khorov, and A.~Lyakhov, ``{On the Joint Usage of
  Target Wake Time and 802.11 ba Wake-Up Radio},'' \emph{IEEE Access}, vol.~8,
  pp. 221\,061--221\,076, 2020.

\bibitem{odelberg20212}
T.~J. Odelberg, J.~Im, and D.~D. Wentzloff, ``{A 2.1 mW- 109dBm NB-IoT Wake-Up
  Receiver},'' in \emph{IEEE Radio Frequency Integrated Circuits Symposium
  (RFIC)}.\hskip 1em plus 0.5em minus 0.4em\relax IEEE, 2021, pp. 235--238.

\bibitem{3gpp}
3GPP, ``{TS 36.211, Section 10.2.6B; Narrowband Wake Up Signal}.''

\bibitem{9626155}
Y.~Mafi, F.~Amirhosseini, S.~A. Hosseini, A.~Azari, M.~Masoudi, and M.~Vaezi,
  ``{Ultra-Low-Power IoT Communications: A novel address decoding approach for
  wake-up receivers},'' \emph{IEEE Transactions on Green Communications and
  Networking}, pp. 1--1, 2021.

\bibitem{heins2022nb}
K.~Heins, ``{NB-IoT Network Deployment},'' in \emph{NB-IoT Use Cases and
  Devices}.\hskip 1em plus 0.5em minus 0.4em\relax Springer, 2022, pp. 81--86.

\bibitem{froytlog2019ultra}
A.~Froytlog, T.~Foss, O.~Bakker, G.~Jevne, M.~A. Haglund, F.~Y. Li, J.~Oller,
  and G.~Y. Li, ``{Ultra-low power wake-up radio for 5G IoT},'' \emph{IEEE
  Communications Magazine}, vol.~57, no.~3, pp. 111--117, 2019.

\bibitem{oller2015has}
J.~Oller, I.~Demirkol, J.~Casademont, J.~Paradells, G.~U. Gamm, and L.~Reindl,
  ``{Has time come to switch from duty-cycled MAC protocols to wake-up radio
  for wireless sensor networks?}'' \emph{IEEE/ACM Transactions on Networking},
  vol.~24, no.~2, pp. 674--687, 2015.

\bibitem{magno2016design}
M.~Magno, V.~Jelicic, B.~Srbinovski, V.~Bilas, E.~Popovici, and L.~Benini,
  ``{Design, implementation, and performance evaluation of a flexible
  low-latency nanowatt wake-up radio receiver},'' \emph{IEEE Transactions on
  Industrial Informatics}, vol.~12, no.~2, pp. 633--644, 2016.

\bibitem{R7C2_1}
K.~Zhou, N.~Nikaein, and T.~Spyropoulos, ``{LTE/LTE-A Discontinuous Reception
  Modeling for Machine Type Communications},'' \emph{IEEE Wireless
  Communications Letters}, vol.~2, no.~1, pp. 102--105, 2013.

\bibitem{ramazanali2016tuning}
H.~Ramazanali and A.~Vinel, ``{Tuning of LTE/LTE-A DRX parameters},'' in
  \emph{IEEE 21st International Workshop on Computer Aided Modelling and Design
  of Communication Links and Networks (CAMAD)}.\hskip 1em plus 0.5em minus
  0.4em\relax IEEE, 2016, pp. 95--100.

\bibitem{R7C2_2}
S.~Huang, G.~Feng, L.~Liang, and S.~Qin, ``{Power-saving coercive sleep mode
  for machine type communications},'' in \emph{23rd Asia-Pacific Conference on
  Communications (APCC)}, 2017, pp. 1--6.

\bibitem{R7C2_3}
H.-L. Chang and M.-H. Tsai, ``{Optimistic DRX for Machine-Type Communications
  in LTE-A Network},'' \emph{IEEE Access}, vol.~6, pp. 9887--9897, 2018.

\bibitem{canweachieve}
J.~Guo, Y.~Chen, J.~Zhu, and S.~Zhang, ``{Can We Achieve Better Wireless
  Traffic Prediction Accuracy?}'' \emph{IEEE Communications Magazine}, vol.~59,
  no.~8, pp. 58--63, 2021.

\bibitem{R7C2_4}
N.~M. Balasubramanya, L.~Lampe, G.~Vos, and S.~Bennett, ``{On Timing
  Reacquisition and Enhanced Primary Synchronization Signal (ePSS) Design for
  Energy Efficient 3GPP LTE MTC},'' \emph{IEEE Transactions on Mobile
  Computing}, vol.~16, no.~8, pp. 2292--2305, 2017.

\bibitem{R7C2_5}
J.~Rinne, J.~Keskinen, P.~R. Berger, D.~Lupo, and M.~Valkama, ``{Viability
  Bounds of M2M Communication Using Energy-Harvesting and Passive Wake-Up
  Radio},'' \emph{IEEE Access}, vol.~5, pp. 27\,868--27\,878, 2017.

\bibitem{rostami2018wireless}
S.~Rostami, K.~Heiska, O.~Puchko, K.~Leppanen, and M.~Valkama, ``{Wireless
  powered wake-up receiver for ultra-low-power devices},'' in \emph{IEEE
  Wireless Communications and Networking Conference (WCNC)}.\hskip 1em plus
  0.5em minus 0.4em\relax IEEE, 2018, pp. 1--5.

\bibitem{rostami2019wake}
S.~Rostami, S.~Lagen, M.~Costa, M.~Valkama, and P.~Dini, ``{Wake-Up Radio Based
  Access in 5G Under Delay Constraints: Modeling and Optimization},''
  \emph{IEEE Transactions on Communications}, vol.~68, no.~2, pp. 1044--1057,
  2019.

\bibitem{eldeeb2021learning}
E.~Eldeeb, M.~Shehab, and H.~Alves, ``{{A Learning-Based Fast Uplink Grant for
  Massive IoT via Support Vector Machines and Long Short-Term Memory}},''
  \emph{IEEE Internet of Things Journal}, 2021.

\bibitem{6629847}
N.~Nikaein, M.~Laner, K.~Zhou, P.~Svoboda, D.~Drajic, M.~Popovic, and S.~Krco,
  ``{Simple Traffic Modeling Framework for Machine Type Communication},'' in
  \emph{The Tenth International Symposium on Wireless Communication Systems},
  2013, pp. 1--5.

\bibitem{lopez2021csi}
O.~L. L{\'o}pez, N.~H. Mahmood, H.~Alves, and M.~Latva-aho, ``{CSI-free vs
  CSI-based multi-antenna WET for massive low-power Internet of Things},''
  \emph{IEEE Transactions on Wireless Communications}, vol.~20, no.~5, pp.
  3078--3094, 2021.

\bibitem{8891507}
O.~Teyeb, A.~Muhammad, G.~Mildh, E.~Dahlman, F.~Barac, and B.~Makki,
  ``{Integrated Access Backhauled Networks},'' in \emph{IEEE 90th Vehicular
  Technology Conference (VTC2019-Fall)}, 2019, pp. 1--5.

\bibitem{abdrabou2021application}
A.~Abdrabou, M.~Al~Darei, M.~Prakash, and W.~Zhuang, ``{Application-Oriented
  Traffic Modeling of WiFi-based Internet of Things Gateways},'' \emph{IEEE
  Internet of Things Journal}, 2021.

\bibitem{madueno2015reliable}
G.~C. Madueno, {\v{C}}.~Stefanovi{\'c}, and P.~Popovski, ``{Reliable and
  efficient access for alarm-initiated and regular M2M traffic in IEEE 802.11
  ah systems},'' \emph{IEEE Internet of Things Journal}, vol.~3, no.~5, pp.
  673--682, 2015.

\bibitem{rostami2019optimized}
S.~Rostami, S.~Lagen, M.~Costa, P.~Dini, and M.~Valkama, ``{Optimized wake-up
  scheme with bounded delay for energy-efficient MTC},'' in \emph{IEEE Global
  Communications Conference (GLOBECOM)}.\hskip 1em plus 0.5em minus 0.4em\relax
  IEEE, 2019, pp. 1--6.

\bibitem{sachs20185g}
J.~Sachs, G.~Wikstrom, T.~Dudda, R.~Baldemair, and K.~Kittichokechai, ``{{5G}
  radio network design for ultra-reliable low-latency communication},''
  \emph{IEEE network}, vol.~32, no.~2, pp. 24--31, 2018.

\bibitem{mahlouji2018analysis}
M.~Mahlouji and T.~Mahmoodi, ``{Analysis of Uplink Scheduling for Haptic
  Communications},'' in \emph{IEEE Globecom Workshops (GC Wkshps)}.\hskip 1em
  plus 0.5em minus 0.4em\relax IEEE, 2018, pp. 1--7.

\bibitem{thomsen2017traffic}
H.~Thomsen, C.~N. Manch{\'o}n, and B.~H. Fleury, ``{A traffic model for
  machine-type communications using spatial point processes},'' in \emph{IEEE
  28th Annual International Symposium on Personal, Indoor, and Mobile Radio
  Communications (PIMRC)}.\hskip 1em plus 0.5em minus 0.4em\relax IEEE, 2017,
  pp. 1--6.

\bibitem{chen2020energy}
Y.~Chen, J.~Yang, X.~Cao, and S.~Zhang, ``{Energy efficiency optimization for
  heterogeneous cellular networks modeled by Mat{\'e}rn hard-core point
  process},'' \emph{China Communications}, vol.~17, no.~8, pp. 70--80, 2020.

\bibitem{deng2020energized}
N.~Deng and M.~Haenggi, ``{The Energized Point Process as a Model for
  Wirelessly Powered Communication Networks},'' \emph{IEEE Transactions on
  Green Communications and Networking}, vol.~4, no.~3, pp. 832--844, 2020.

\bibitem{li2020resource}
F.~Li and K.-Y. Lam, ``{Resource optimization in satellite-based Internet of
  Things},'' in \emph{International Conference on Artificial Intelligence in
  Information and Communication (ICAIIC)}.\hskip 1em plus 0.5em minus
  0.4em\relax IEEE, 2020, pp. 006--011.

\bibitem{zhang2020opportunities}
X.~Zhang, J.~Grajal, M.~L{\'o}pez-Vallejo, E.~McVay, and T.~Palacios,
  ``{Opportunities and challenges of ambient radio-frequency energy
  harvesting},'' \emph{Joule}, vol.~4, no.~6, pp. 1148--1152, 2020.

\bibitem{graph}
K.~He, X.~Chen, Q.~Wu, S.~Yu, and Z.~Zhou, ``{Graph Attention Spatial-Temporal
  Network With Collaborative Global-Local Learning for Citywide Mobile Traffic
  Prediction},'' \emph{IEEE Transactions on Mobile Computing}, vol.~21, no.~4,
  pp. 1244--1256, 2022.

\bibitem{hybrid_deep}
S.~Mahajan, H.~R., and K.~Kotecha, ``{Prediction of Network Traffic in Wireless
  Mesh Networks Using Hybrid Deep Learning Model},'' \emph{IEEE Access},
  vol.~10, pp. 7003--7015, 2022.

\bibitem{dmTP}
Z.~Zhang, F.~Li, X.~Chu, Y.~Fang, and J.~Zhang, ``{dmTP: A Deep Meta-Learning
  Based Framework for Mobile Traffic Prediction},'' \emph{IEEE Wireless
  Communications}, vol.~28, no.~5, pp. 110--117, 2021.

\bibitem{wang2017spatiotemporal}
J.~Wang, J.~Tang, Z.~Xu, Y.~Wang, G.~Xue, X.~Zhang, and D.~Yang,
  ``{Spatiotemporal modeling and prediction in cellular networks: A big data
  enabled deep learning approach},'' in \emph{IEEE INFOCOM - IEEE Conference on
  Computer Communications}.\hskip 1em plus 0.5em minus 0.4em\relax IEEE, 2017,
  pp. 1--9.

\bibitem{memon2019artificial}
M.~L. Memon, M.~K. Maheshwari, N.~Saxena, A.~Roy, and D.~R. Shin, ``{Artificial
  intelligence-based discontinuous reception for energy saving in 5G
  networks},'' \emph{Electronics}, vol.~8, no.~7, p. 778, 2019.

\bibitem{bellman2015applied}
R.~E. Bellman and S.~E. Dreyfus, \emph{{Applied dynamic programming}}.\hskip
  1em plus 0.5em minus 0.4em\relax Princeton university press, 2015.

\bibitem{luong2019applications}
N.~C. Luong, D.~T. Hoang, S.~Gong, D.~Niyato, P.~Wang, Y.-C. Liang, and D.~I.
  Kim, ``{Applications of deep reinforcement learning in communications and
  networking: A survey},'' \emph{IEEE Communications Surveys \& Tutorials},
  vol.~21, no.~4, pp. 3133--3174, 2019.

\bibitem{xu2021generative}
Y.-H. Xu, X.~Liu, W.~Zhou, and G.~Yu, ``{Generative Adversarial LSTM Networks
  Learning for Resource Allocation in UAV-served M2M Communications},''
  \emph{IEEE Wireless Communications Letters}, 2021.

\bibitem{CRAN}
Y.~Fu and X.~Wang, ``{Traffic Prediction-Enabled Energy-Efficient Dynamic
  Computing Resource Allocation in CRAN Based on Deep Learning},'' \emph{IEEE
  Open Journal of the Communications Society}, vol.~3, pp. 159--175, 2022.

\bibitem{gers2000learning}
F.~A. Gers, J.~Schmidhuber, and F.~Cummins, ``{Learning to forget: Continual
  prediction with LSTM},'' \emph{Neural computation}, vol.~12, no.~10, pp.
  2451--2471, 2000.

\bibitem{eramo2022application}
V.~Eramo and T.~Catena, ``{Application of an Innovative Convolutional/LSTM
  Neural Network for Computing Resource Allocation in NFV Network
  Architectures},'' \emph{IEEE Transactions on Network and Service Management},
  2022.

\bibitem{chen2021expressway}
Z.~Chen, B.~Wu, B.~Li, and H.~Ruan, ``{Expressway Exit Traffic Flow Prediction
  for ETC and MTC Charging System Based on Entry Traffic Flows and LSTM
  Model},'' \emph{IEEE Access}, vol.~9, pp. 54\,613--54\,624, 2021.

\bibitem{sharma2017adam}
M.~Sharma, R.~Pachori, and A.~Rajendra, ``{Adam: a method for stochastic
  optimization},'' \emph{Pattern Recogn. Lett}, vol.~94, pp. 172--179, 2017.

\bibitem{MATLAB}
\emph{{MATLAB version 9.7.0.1190202 (R2018b)}}, The Mathworks, Inc., Natick,
  Massachusetts, 2018.

\end{thebibliography}

  \vspace{1em}

\begin{wrapfigure}{l}{25mm} 
    \includegraphics[width=1in,height=1.25in,clip,keepaspectratio]{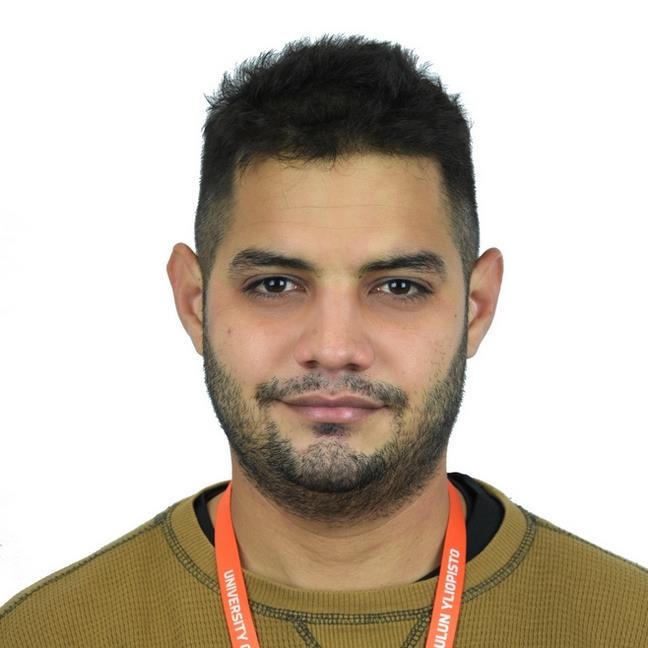}
\end{wrapfigure}\par
  \noindent \textbf{David E. Ruíz-Guirola} %was born in Santa Clara, Cuba, in 1994. He 
  received the B.Sc. (1st class honors, 2018) and M.Sc. (with distinction, 2019) degree in Telecommunications and Electronic Engineering from the Central University of Las Villas (UCLV), Santa Clara, Cuba. From 2018-2021 he served as an Associate Professor %with the Department of Telecommunications, 
  at UCLV. He joined the Centre for Wireless Communications, University of Oulu, Finland, in 2021. He is currently pursuing the Ph.D. degree at the University of Oulu. His research interests include sustainable IoT, energy harvesting, wireless RF energy transfer, machine-type communications, machine learning, and traffic prediction.\par
    
  \vspace{1em}
  
  \begin{wrapfigure}{l}{25mm} 
    \includegraphics[width=1in,height=1.25in,clip,keepaspectratio]{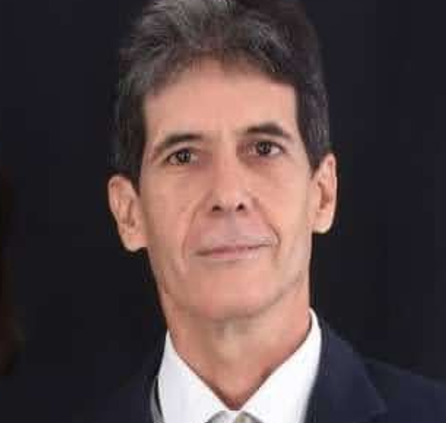}
\end{wrapfigure}\par
  \noindent \textbf{Carlos A. Rodríguez-López} %was born in Cienfuegos, Cuba, in 1966. He 
  received the B.Sc. and M.Sc. degrees in Electronic Equipment and Components and Telecommunications from the Central University of Las Villas (UCLV), Cuba, in 1990 and 2000, respectively. He is currently a Professor with the Department of Telecommunications, UCLV. His research interests are in the area of wireless mobile communications, including visible light communications.\par
    
  \vspace{1em}
  
  \begin{wrapfigure}{l}{25mm} 
    \includegraphics[width=1in,height=1.25in,clip,keepaspectratio]{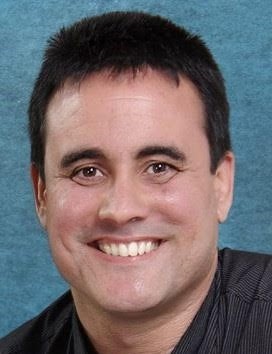}
\end{wrapfigure}\par
  \noindent \textbf{Samuel Montejo-Sánchez} (IEEE M'17-SM'22) %was born in Camagüey, Cuba, in 1979. He 
  received the B.Sc., M.Sc., and D.Sc. degrees in telecommunications from the Central University of Las Villas (UCLV), Cuba, in 2003, 2007 and 2013, respectively. From 2003-2017, he was an Associate Professor %with the Department of Telecommunications, 
  at UCLV. %In 2011, he was Visiting Ph.D. Student with the Federal University of Technology (UTFPR), Brazil. In 2017, he held a postdoctoral position with the University of Chile. 
  Since 2018, he has been with the Programa Institucional de Fomento a la I+D+i (PIDi), Universidad Tecnológica Metropolitana (UTEM). He leads the FONDECYT Iniciación No. 11200659 (Toward High Performance Wireless Connectivity for IoT and Beyond-5G Networks) project. %and is member of the FONDECYT Regular No. 1201893 (IoT goes to Space: Wireless Networking protocols and architectures for IoT networks served by LEO satellite constellations) and FONDEQUIP EQM180180 (Clúster Supermicro para Cómputo Científico) projects. 
  His research interests include wireless communications, 
  %cognitive radio, network coding, energy efficiency, vehicular ad hoc networks, physical layer security, and the Internet of Things. 
  signal processing, sustainable IoT, and wireless RF energy transfer.
  He was a co-recipient of the 2016 Research Award from the Cuban Academy of Sciences.\par
    
  \vspace{1em}
  
  \begin{wrapfigure}{l}{25mm} 
    \includegraphics[width=1in,height=1.25in,clip,keepaspectratio]{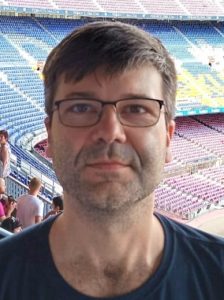}
\end{wrapfigure}\par
  \noindent \textbf{Richard Demo Souza}  (IEEE S'01–M'04–SM'12) received the D.Sc. degree in Electrical Engineering from the Federal University of Santa Catarina (UFSC), Brazil, in 2003. From 2004 to 2016 he was with the Federal University of Technology – Paraná (UTFPR), Brazil. Since 2017 he has been with UFSC, where he is a Professor. His research interests are in the areas of wireless communications and signal processing. He has served as Editor or Associate Editor for the SBrT Journal of Communications and Information Systems, the IEEE Communications Letters, the IEEE Transactions on Vehicular Technology, the IEEE Transactions on Communications, and the IEEE IoT Journal. He is a co-recipient of the 2014 IEEE/IFIP Wireless Days Conference Best Paper Award, the supervisor of the awarded Best PhD Thesis in Electrical Engineering in Brazil in 2014, and a co-recipient of the 2016 Research Award from the Cuban Academy of Sciences.\par
  
  \vspace{1em}
  
  \begin{wrapfigure}{l}{25mm} 
    \includegraphics[width=1in,height=1.25in,clip,keepaspectratio]{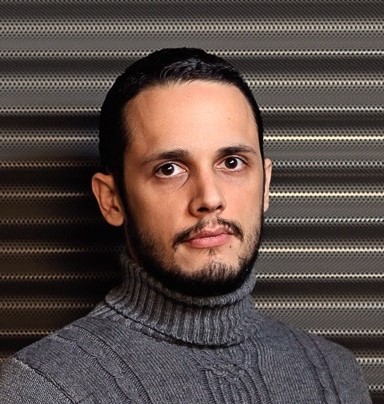}
    \end{wrapfigure}\par
  \noindent \textbf{Onel L. A. López} (S'17-M'20) received the B.Sc. (1st class honors, 2013), M.Sc. (2017) and D.Sc. (with distinction, 2020) degree in Electrical Engineering from the Central University of Las Villas (Cuba),  the Federal University of Paraná (Brazil), and the University of Oulu (Finland), respectively. 
  %From 2013-2015 he served as a specialist in telematics at the Cuban telecommunications company (ETECSA). 
  He is a collaborator to the 2016 Research Award given by the Cuban Academy of Sciences, a co-recipient of the 2019 IEEE EuCNC Best Student Paper Award, and the recipient of the 2020 best doctoral thesis award granted by Finland TEK and TFiF in 2021. He is co-author of the book entitled ``Wireless RF Energy Transfer in the massive IoT era: towards sustainable zero-energy networks'', Wiley, Dec 2021. He currently holds an Assistant Professorship (tenure track) in sustainable wireless communications engineering in the Centre for Wireless Communications (CWC), Oulu, Finland. His research interests include wireless communications, signal processing, sustainable IoT, and wireless RF energy transfer.
  \par
  
  \vspace{2em}
  
  \begin{wrapfigure}{l}{25mm} 
    \includegraphics[width=1in,height=1.25in,clip,keepaspectratio]{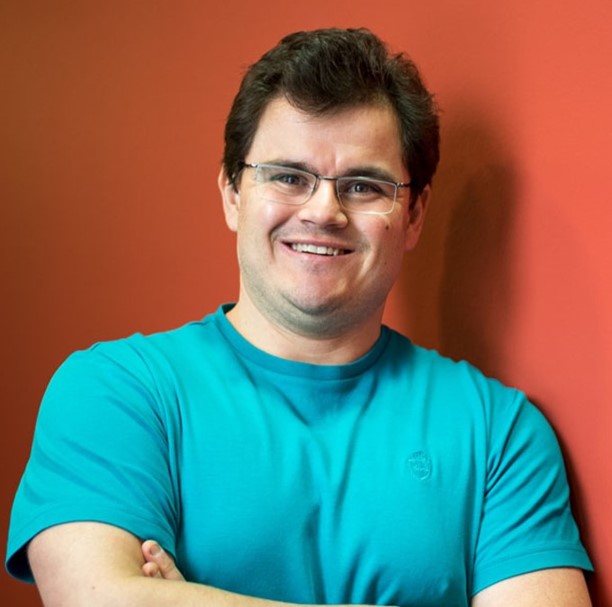}
\end{wrapfigure}\par
  \noindent \textbf{Hirley Alves} (S'11–M'15) is Associate Professor and Head of the Machine-type Wireless Communications Group at the 6G Flagship, Centre for Wireless Communications, University of Oulu. He is actively working on massive connectivity and ultra-reliable low latency communications for future wireless networks, 5GB and 6G, full-duplex communications, and physical-layer security. In addition, he leads the URLLC activities for the 6G Flagship Program. He has received several awards and has been the organizer, chair, TPC, and tutorial lecturer for several renowned international conferences. He is the General Chair of the ISWCS'2019 and the General Co-Chair of the 1st 6G Summit, Levi 2019, and ISWCS 2021.\par

\end{document}